\documentclass[nofootinbib,floats,letterpaper,floatfix,eqsecnum,prd]{revtex4}

\usepackage{dcolumn,epsfig}
\usepackage{amssymb,amsmath}
\usepackage[usenames]{color}

\def\be{\begin{equation}}
\def\ee{\end{equation}}
\def\beq{\begin{eqnarray}}
\def\eeq{\end{eqnarray}}

\def\IL{\relax{\rm I\kern-.18em L}}

\def\f{\frac}

\def\h{\eta}

\def\m{\mu}
\def\n{\nu}
\def\r{\rho}
\def\s{\sigma}

\def\p{\partial}

\def\f{\varphi}
\def\a{\alpha}
\def\b{\beta}

\def\l{\lambda}
\def\g{\gamma}

\def\D{\Delta}
\def\w{\omega}

\def\nb{\nonumber}

\def\ba{\begin{eqnarray}}
\def\ea{\end{eqnarray}}

\begin{document}

\title{Phase transitions between   Reissner-Nordstrom and dilatonic
black holes in 4D AdS spacetime}

\author{Mariano Cadoni}\email{email: mariano.cadoni@ca.infn.it }
\affiliation{Dipartimento di Fisica, Universit\`a di Cagliari and
INFN, Sezione di Cagliari }
\author{Giuseppe D'Appollonio}\email{email: giuseppe.dappollonio@ca.infn.it }
\affiliation{Dipartimento di Fisica, Universit\`a di Cagliari and
INFN, Sezione di Cagliari }
\author{Paolo Pani}\email{email: paolo.pani@ca.infn.it }
\affiliation{Dipartimento di Fisica, Universit\`a di Cagliari and
INFN, Sezione di Cagliari }
\begin{abstract}
We study Einstein-Maxwell-dilaton gravity models in four-dimensional
anti-de Sitter (AdS) spacetime which admit the Reissner-Nordstrom
(RN) black hole solution. We show that below a critical temperature
the AdS-RN solution becomes unstable against scalar perturbations
and the gravitational system undergoes a phase transition. We show
using numerical calculations that the new phase is a charged
dilatonic black hole. Using the AdS/CFT correspondence we discuss
the phase transition in the dual field theory both for non-vanishing
temperatures and in the extremal limit. The extremal solution has a
Lifshitz scaling symmetry. We discuss the optical conductivity in
the new dual phase and find interesting behavior at low frequencies
where it shows a ``Drude peak''. The resistivity varies with
temperature in a non-monotonic way and displays a minimum at low
temperatures which is reminiscent of the celebrated Kondo effect.

\end{abstract}
\maketitle



\section{Introduction}

The AdS/CFT correspondence \cite{Maldacena:1997re} provides a
deep connection between quantum gravity
and quantum gauge theories.
When the classical gravity approximation is reliable, it also provides efficient techniques
for the computation of the thermodynamical and transport properties
of strongly interacting quantum field theories.
This holographic approach
has been applied both to gauge theories similar to QCD
\cite{Son:2007vk, Mateos:2007ay} and, more recently, to
condensed matter phenomena
\cite{Sachdev:2008ba, Herzog:2009xv, Hartnoll:2009sz, McGreevy:2009xe}.
It relies on the identification between the black hole solutions of the bulk theory
and the thermal states of the boundary theory.
Transport coefficients can be computed at strong coupling
by solving the equations governing small perturbations of the black hole
background \cite{Son:2002sd}.

In the attempt of developing a gravitational description for
condensed matter phenomena one is primarily interested in finding
black hole solutions that can capture the salient features of
realistic many-body systems and in studying the corresponding phase
diagram. The existence and the stability properties of a given
solution depend on the field content and on the couplings of the
bulk action. As an illustration of this point let us consider the
simplest example, relevant for the holographic description of
systems at finite temperature and charge density. The minimal bulk
theory is in this case Einstein-Maxwell theory and the only static
charged black hole solution is the AdS-Reissner-Nordstr\"om (AdS-RN)
black hole \cite{Israel:1967wq}. The solution remains stable as we
lower the temperature and the ground state of the system corresponds
to the extremal AdS-RN black hole. If we include in the action a
charged scalar field with a minimal coupling to the gauge field,
below a critical temperature new branches of solutions appear which
are thermodynamically favoured. The resulting instability of the
AdS-RN black hole provides a holographic description of a phase
transition in the dual theory \cite{Gubser:2008wv,
Hartnoll:2008kx,Hartnoll:2008vx}. Below the critical temperature the
new solution is a static charged black hole with a scalar hair. The
non-trivial profile of the scalar field corresponds to a charged
condensate in the dual theory that breaks spontaneously a global
$U(1)$ symmetry. One then expects to observe in the new phase
phenomena typical of superfluid or superconducting systems and this
is confirmed by the study of the linearized response of the hairy
black hole to small perturbations
\cite{Hartnoll:2008kx,Hartnoll:2008vx}.

Most investigations considered so far only the case of scalar fields
minimally coupled to the electromagnetic field. Relatively little is
known about the stability of the AdS-RN solution in non-minimally
coupled models (Einstein-Maxwell-dilaton gravity). Non-minimal
couplings of the form $f(\phi) F^{2}$ between a scalar fields $\phi$
and the Maxwell tensor  are very common in supergravity and in the
low-energy effective action of string theory models. In flat
spacetime charged black holes solutions of Einstein-Maxwell-dilaton
gravity are well-known
\cite{Gibbons:1987ps,Garfinkle:1990qj,Monni:1995vu}. These solutions
involve a non-constant scalar field and differ significantly from
the RN black hole since there is no inner horizon and the event
horizon becomes singular in the extremal limit. Examples of charged
dilaton black hole solutions with AdS asymptotics are provided by
the family of four-charge black holes in ${\cal N}=8$
four-dimensional gauged supergravity \cite{Duff:1999gh}. As shown in
\cite{Gubser:2009qt} in the extremal limit  these black holes can
support an isolated fermion normal mode which signals the presence
of a Fermi surface in the dual systems. Quite interestingly the
AdS-RN black hole is not a solution of dilaton gravity models with
$f=\exp(a\phi)$.

In this paper we shall consider models where the coupling between
the scalar field and the kinetic term of the gauge field starts
quadratically. We first investigate the stability  of the AdS-RN
black hole under scalar perturbations and show that it becomes
unstable below a critical temperature. In flat space this kind of
instability was studied in \cite{Gubser:2005ih}. In
\cite{Gubser:2000ec} it was found that the AdS-RN black hole in
${\cal N}=8$ four-dimensional gauged supergravity is unstable
against fluctuations involving a scalar and four gauge fields. In
\cite{Hartnoll:2008hs} a similar instability was found studying
scalar fluctuations around a dyonic black hole in order to compute
the momentum relaxation time scale induced by the presence of
impurities.

In our models AdS-RN and dilaton black holes can coexist. Below a
critical temperature the AdS-RN solution undergoes a phase
transition toward a new solution, which is a hairy black hole
solution. Following closely the approach of
\cite{Hartnoll:2008kx,Hartnoll:2008vx} we will construct numerically
the hairy black hole solutions. In order to clarify the behavior of
the model at low temperatures we will also study the extremal
solution. The near horizon form is characterized by a Lifshitz
scaling isometry \cite{Kachru:2008yh} and  is a simple
generalization of the solution found in \cite{Goldstein:2009cv}.

In the new phase a neutral scalar operator  acquires  a
non-vanishing expectation value. The condensate modifies the
transport properties of the system in an interesting way. This is
clearly illustrated by the behavior  we find for  the optical
conductivity at small frequencies and non-vanishing temperatures:
there is a minimum and then the conductivity increases until it
reaches a value which can be much higher than the constant value at
high frequencies. We can understand this fact by rewriting the
equation for the vector fluctuations as a Schr\"odinger equation
\cite{Horowitz:2009ij}. The non-minimal coupling between the scalar
and the gauge field induces a term in the potential that is not
positive definite. A similar behavior was recently observed in
models where the Born-Infeld action of a probe brane is coupled to a
geometry with a Lifshitz scaling symmetry \cite{Hartnoll:2009ns}.

Another interesting feature of the new phase is the fact that the
resistivity does not increase monotonically with the temperature but
displays a minimum. A  similar behavior of the  resistivity is
observed in metals containing magnetic impurities, an effect
explained by Kondo as resulting from the interaction between the
magnetic moment of the conduction electrons and the impurity.

The plan of this paper is as follows. In Section
\ref{sec:instability} we present our model and discuss the
stability of the AdS-RN black hole against scalar perturbations,
providing approximate criteria to identify the region in parameter space where an instability
is likely to occur. In Section \ref{sec:backreaction} we construct
numerically the charged black
hole solutions with a neutral scalar hair and show that they are
thermodynamically favoured. In Sec. \ref{sec:Tcon0} we give the
analytic form of the near-horizon solution for the extremal and near-extremal
black holes. In Section \ref{sec:holography} we analyze the optical conductivity
in the new phase of the dual field theory described by the dilaton black hole.
In Sec. \ref{sec:conclusions} we present our conclusions.


\section{Instability of AdS-RN black holes in
Einstein-Maxwell-dilaton gravity}\label{sec:instability}

In this paper we consider Einstein gravity coupled to an abelian
gauge field $A_\m$ and a real scalar field $\phi$. The Lagrangian
\be {\cal L}=R-\frac{f(\phi)
}{4}F^2-\frac{1}{2}\partial^\mu\phi\partial_\mu\phi-V(\phi)\,,\label{lagr_gen}
\ee
depends on the choice of two functions, a potential $V(\phi)$ and a
function $f(\phi)$ that couples the scalar to the kinetic term
of the gauge field. Lagrangians of this type are common in
supergravity and in the low-energy limit of string theory models.
The equations of motion read
\beq\label{max_scal_eq} &&\nabla_\mu\left(f(\phi)F^{\mu\nu}\right)=0
\,, \nb \\
&&\nabla^2\phi=\frac{dV(\phi)}{d\phi}+\frac{df(\phi)}{d\phi}\frac{F^2}{4}\,,  \\
&&R_{\mu\nu} - \frac{1}{2} g_{\mu \nu} R  = -
\frac{f(\phi)}{2}\left( F_{\mu \rho} F^\rho{}_\n + \frac{g_{\mu
\nu}}{4} F^{\rho\sigma} F_{\rho\sigma}\right)  + \frac{1}{2} \left (
\partial_{\mu}\phi\partial_\nu \phi - \frac{g_{\m\n}}{2} \partial^{\rho}
\phi\partial_{\rho}\phi \right ) - \frac{g_{\m\n}}{2} V(\phi) \,. \nb
\label{einsteineq} \eeq

We shall restrict the possible choices of $V(\phi)$ and $f(\phi)$ by
imposing two requirements. The first is that the potential $V(\phi)$
admits stable AdS vacua. Assuming for simplicity that there is only
one extremum at $\phi = 0$, the potential can be expanded for small
values of the field as \be V(\phi)=-\frac{6}{L^2}+\frac{\b}{2
L^2}\phi^2+{\cal O}(\phi^3)\,, \ee where $L$ is the AdS radius and
$\b$ parametrizes the mass of the field, $m^2 L^2 = \b$. The AdS
vacuum is stable if the mass parameter satisfies the
Breitenlohner-Freedman (BF) bound $\b \ge - 9/4$
\cite{Breitenlohner:1982bm}. In the following we will limit our
discussion to quadratic potentials and to potentials of the form
$V(\phi) = - 2W_0 \cosh (b \phi)$. In the latter case $L^2 = 3/W_0$,
$\b = - 6 b^2$ and the BF bound becomes $b^2 \le 3/8$.

The second requirement is that the AdS-RN black hole is a solution
of the equations (\ref{max_scal_eq}). This is the case if
the first derivative of the coupling function vanishes at the
extremum of $V$, $\frac{df}{d\phi}(0) = 0$. For small values of the field
the function $f$ can be expanded as \be
f(\phi)=1+\frac{\alpha}{2}\phi^2+{\cal O}(\phi^3)\, ,
\label{expVandf} \ee where the parameter $\a$ is assumed to be non-negative.
In the following we will consider mainly quadratic
coupling functions and functions of the form $f(\phi) = \cosh (a
\phi)$ for which $\alpha = a^2$.
In our analysis of the zero temperature limit of the
hairy black hole solutions in Section \ref{sec:Tcon0}
we will also consider exponential coupling functions.

We will look for static electrically charged solutions of the
equations of motion with translational symmetry in two spatial directions.
The metric can be written
as \be
ds^2=-g(r)e^{-\chi(r)}dt^2+\frac{dr^2}{g(r)}+r^2(dx^2+dy^2)\,.\label{metric_ansatz_BR}
\ee
The scalar field is $\phi=\phi(r)$ and only the temporal component of the gauge potential
is non-vanishing, $A_0=A_0(r)$. The equations
of motion become
\beq
\phi''+\left(\frac{g'}{g}-\frac{\chi'}{2}+\frac{2}{r}\right)\phi'(r)-\frac{1}{g}\frac{d
V}{d \phi}+\frac{{A_0'}^2e^\chi}{2g} \frac{d f}{d \phi}&=&0\,,\label{eq:BR_scalar}\\
(r^2 e^{\frac{\chi}{2}} f(\phi) A_0')'  &=& 0 \,,\label{eq:BR_maxwell}\\
\chi'+\frac{r{\phi'}^2}{2}&=&0 \label{eq:BR_einstein0}\,,\\
\frac{{\phi'}^2}{4}+\frac{{A_0'}^2e^\chi
f(\phi)}{4g}+\frac{g'}{rg}+\frac{1}{r^2}+\frac{V(\phi)}{2g}&=&0\, ,
\label{eq:BR_einstein1} \eeq 
where here and in the following a prime will always denote a derivative with respect to $r$.
When the
condition ($\ref{expVandf}$) is satisfied, the equations of motion
admit the AdS-RN black hole solution \be
g=-\frac{2M}{r}+\frac{Q^2}{4r^2}+\frac{r^2}{L^2}\,,\qquad \chi = 0
 \,,\qquad
A_0=\frac{Q}{r}-\frac{Q}{r_h}\,, \qquad \phi = 0 \,, \label{eq:RSAdS} \ee
where $M$ and $Q$ are respectively the mass and the electric charge
of the black hole and $r_h$ the radius of the horizon. The black
hole temperature is \be 4 \pi T_{RN} =  \frac{3 r_h}{L^2} -
\frac{Q^2}{4 r_h^3} \ , \label{trn} \ee and the solution becomes
extremal for $12 r_h^4= Q^2 L^2$.

Let us now discuss the stability of the AdS-RN solution
(\ref{eq:RSAdS}) against small perturbations of the scalar field.
Given the planar symmetry of the solution, we can expand the scalar
perturbation in Fourier modes  \be \phi_{\w, \vec{k}} =
\frac{R(r)}{r} e^{i (k_1 x + k_2 y - \omega t)} \ . \ee The radial
function $R$ solves the following equation
\be g^2 R''+g g'R' +\left[\omega^2-V
\right]R=0\,,\label{scalar_pert} \hspace{1cm} V = g
\left[\frac{\vec{k}^2}{r^2} + \frac{g'}{r} +
m^2_{\text{eff}}\right]\, ,  \ee
where we have introduced an effective mass
\be m^2_{\text{eff}}(r) = m^2-\frac{\alpha}{2} {A_0'(r)}^2 \ .
\label{eff_mass} \ee
In the presence of an electric background field the non-minimal
coupling gives a negative contribution to the effective mass. If the
coupling is strong enough it can lower the mass below the BF bound
and destabilize the background. This mechanism of instability can be
generalized to magnetic and dyonic black holes. In this case whether
the solution is stable or unstable depends on the sign of the coupling
$\alpha$ and on the relative magnitude of the electric and magnetic charges.
In flat space this kind of instability was studied in \cite{Gubser:2005ih}. In \cite{Gubser:2000ec}
a dynamical instability of the AdS-RN black hole in ${\cal N}=8$ four-dimensional
gauged supergravity was found involving fluctuations of both scalar and 
gauge fields. A similar instability was also found in \cite{Hartnoll:2008hs} analysing
scalar fluctuations around a dyonic black hole in order to compute the
momentum relaxation time scale induced by the presence of impurities.

Before studying numerically the solutions of Eq.~$(\ref{scalar_pert})$, it is worth discussing some approximate
criteria that point to the instability of the AdS-RN background.
When $\b = - 2$ we can in fact provide a simple proof of the
instability. In this case the potential $V(r)$ vanishes both at the
horizon and at infinity and a sufficient condition for the existence
of unstable modes is given by \cite{Dotti:2004sh}
\be \int_{r_h}^\infty dr{\frac{V(r)}{g(r)}}<0\,. \ee
In terms of the parameter $\a$ this instability condition becomes 
\be \alpha > \frac{3r_h^4}{Q^2L^2} -\frac{1}{4} \,. \label{critical_alpha}\ee  
For generic values of $\b$ an approximate condition for the
instability of the solution follows from the observation that the
region where the effective mass is below the BF bound
contributes to the formation of a tachyonic mode whereas the region
where the effective mass is above the BF bound supports stability.
We can characterize the two regions introducing an  instability radius
$r_i$
\be r^4_i = \frac{Q^2 L^2}{\g}\,,\qquad
\gamma=\frac{2}{\alpha}\left({\frac{9}{4}+\beta}\right)\, .
\label{rc} \ee
%
The region where $m^2_{\rm eff} < -9/4$ corresponds to $r<r_i$ while
the region where $m^2_{\rm eff} > -9/4$ corresponds to $r>r_i$. When
$r_i\gg r_h$ the AdS-RN black hole is likely to be unstable and since $r_i\to\infty$
when $\a \to \infty$ or when $\beta \to -9/4$ we expect an
instability for large values of $\a$ or for masses close to the BF
bound. This approximate condition can be expressed in terms of the
black hole temperature $(\ref{trn})$ as \be T_{i} \gg T_{RN} \ ,
\hspace{1cm} T_i = \frac{\sqrt{QL}}{16\pi L^2}\left(\frac{12 -
\g}{\g^{1/4}}\right) \label{Tc} \,. \ee

\begin{figure}[ht]
\begin{center}
\epsfig{file=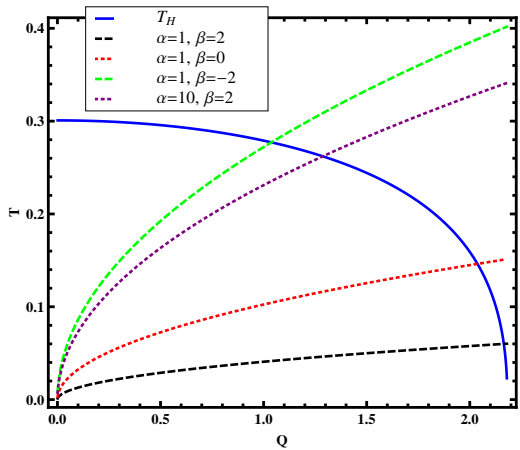,width=7cm,angle=0} \caption{The
instability temperature $T_i$ and the black hole temperature $T_{RN}$
as functions of $Q$ for $L=M=1$. Plots extend from $Q=0$ to
$Q_\text{ext}=M\sqrt{3}(2L/M)^{1/3}$ which corresponds to extremal
black holes. The region of instability is $T_{RN} \ll T_i$. Curves
are ordered in a counterclockwise sense for decreasing values of
$\gamma$. The smaller $\gamma$ the larger the instability region. }
\label{fig:Tc}
\end{center}
\end{figure}

The instability temperature is shown in  Figure \ref{fig:Tc} as a
function of the black hole charge $Q$. The region where an instability is
expected to occur is
$T_{RN}\ll T_i$, i.e. $Q\gg Q_i$, where $Q_i$ is given by the
intersection point of the two curves $T_{RN}(Q)$ and $T_i(Q)$.

%
%

\begin{figure}[ht]
\begin{center}
\begin{tabular}{ccc}
\epsfig{file=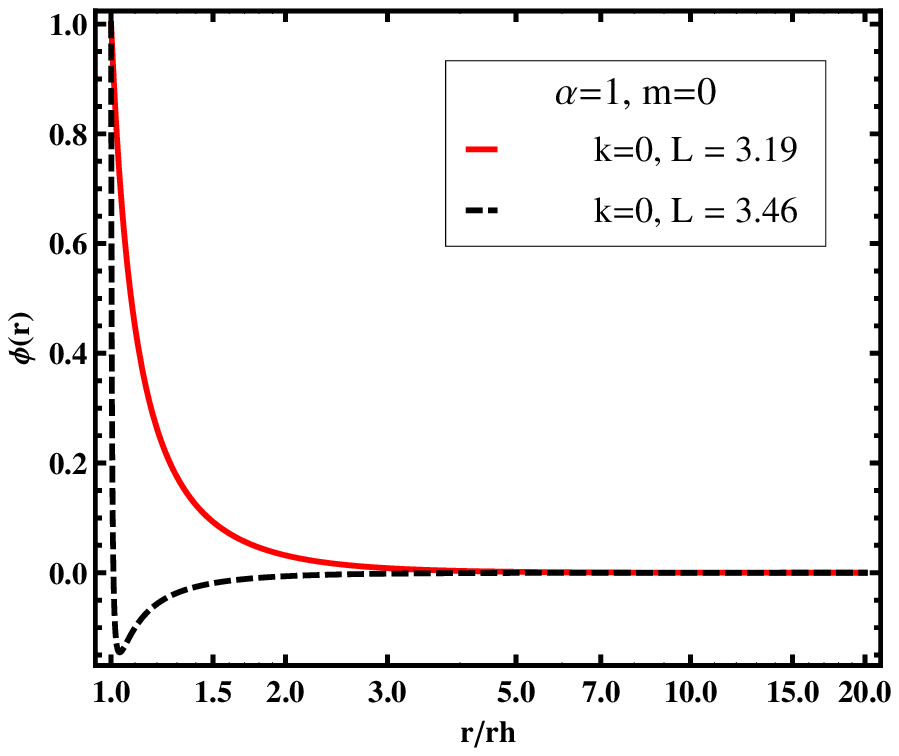,width=5cm,angle=0}&
\epsfig{file=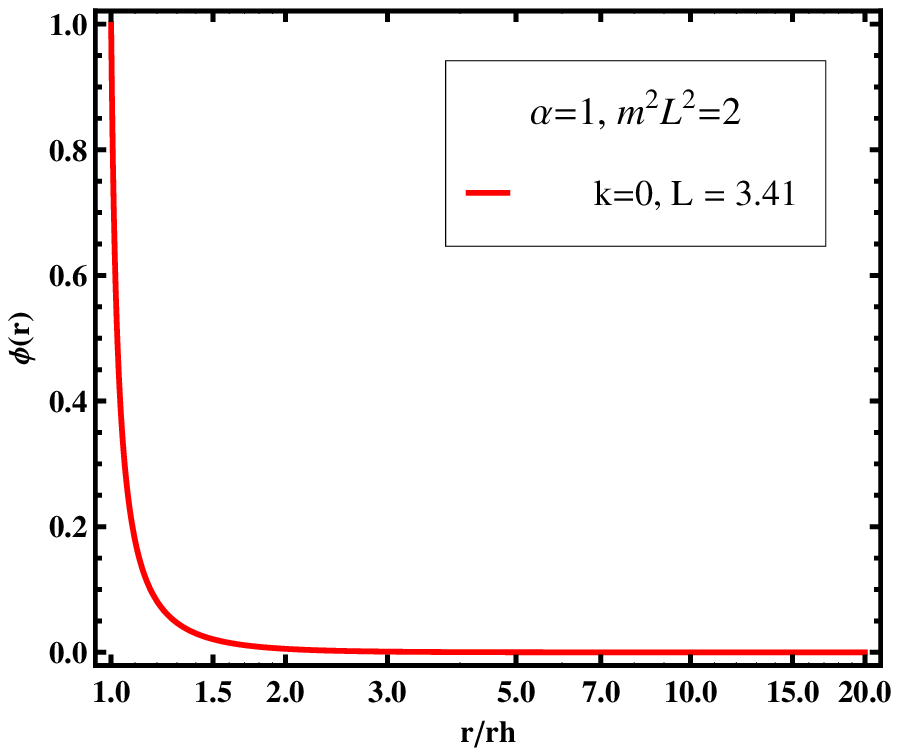,width=5cm,angle=0}&
\epsfig{file=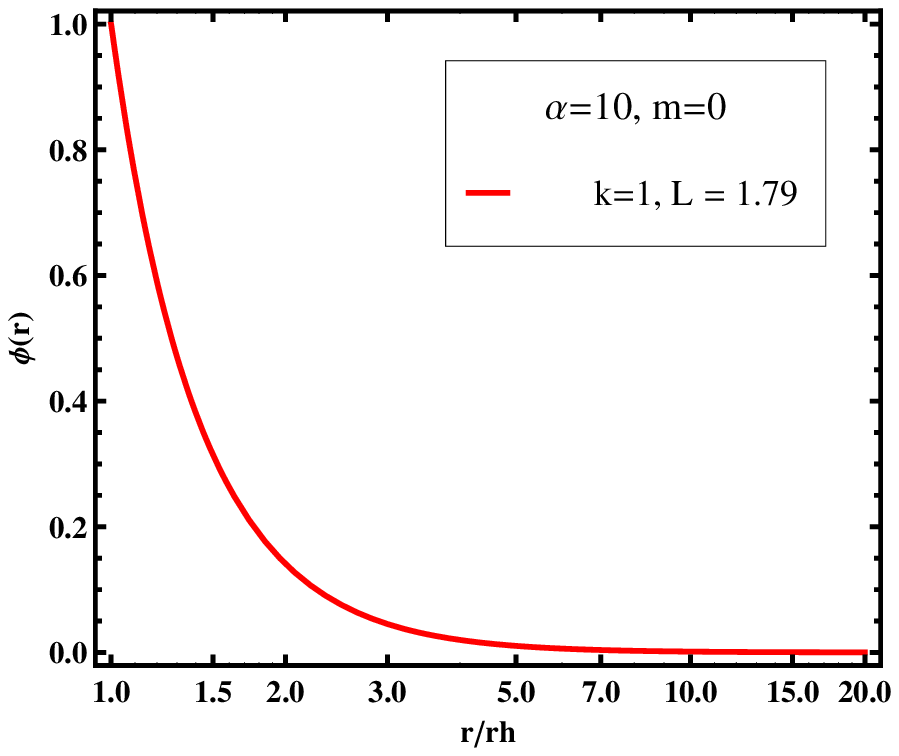,width=5cm,angle=0}
\end{tabular}
\caption{Some examples of marginally stable modes with $m^2 \ge 0$
\label{fig:modes10}}
\end{center}
\end{figure}

\begin{figure}[ht]
\begin{center}
\begin{tabular}{cc}
\epsfig{file=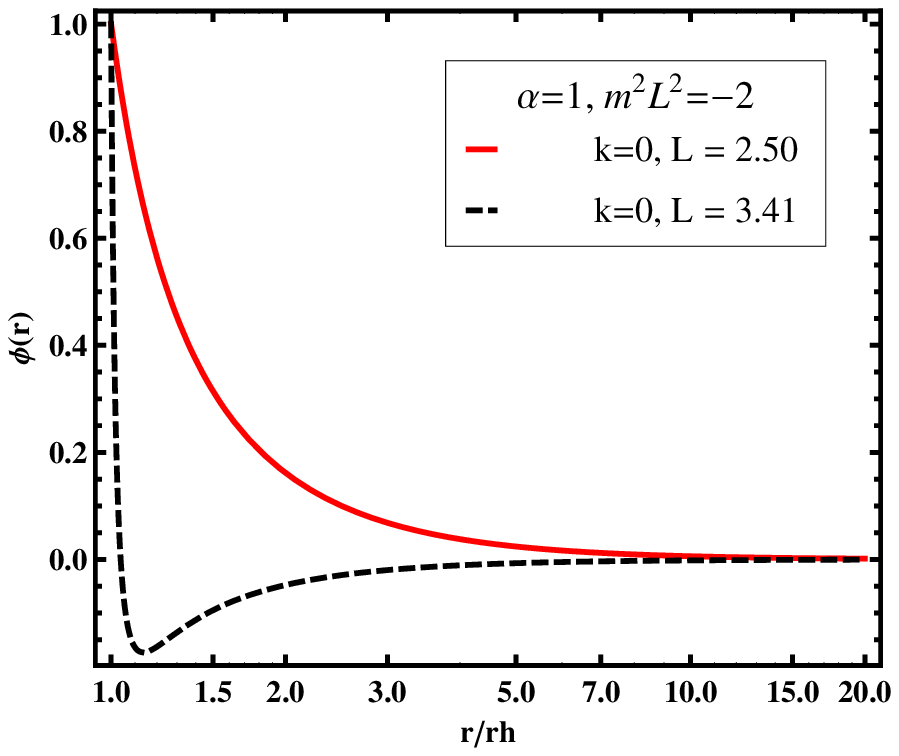,width=5cm,angle=0}&
\epsfig{file=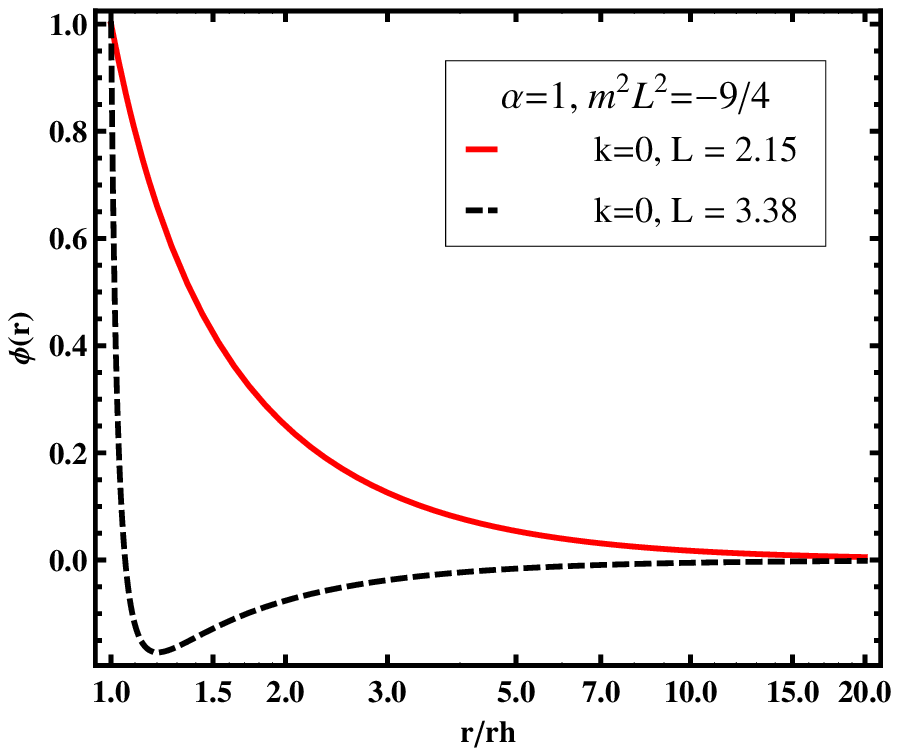,width=5cm,angle=0}
\end{tabular}
\caption{Some examples of marginally stable modes with $0 \ge
m^2\geq m_{BF}^2$\label{fig:modes20}}
\end{center}
\end{figure}

In order to confirm the existence of the instability one should find
unstable modes of the scalar equation $(\ref{scalar_pert})$.
Unstable modes correspond to normalizable solutions of $(\ref{scalar_pert})$ with
purely ingoing boundary conditions at the horizon and complex
frequency $\w$ with $\text{Im}(\omega)>0$ (see \cite{Berti:2009kk}
for a recent review on black hole perturbations). These modes grow
exponentially in time destabilizing the background. Another strong
indication of the instability of a gravitational background is
provided by the presence of marginally stable modes, namely modes
with $\w = 0$. Setting also $\vec{k} = 0$, Eq.~(\ref{scalar_pert})
reduces to
\be \left(\square-m^2_{\text{eff}}\right)\phi(r)=0\, .
\label{marginal} \ee
We can solve the previous equation by numerical integration starting
from a series expansion at the horizon and imposing suitable
boundary conditions near the AdS boundary.
The expansion at large $r$ of the scalar field is
\be \phi\sim \frac{{\cal O}_{\D}}{r^{\D}} \ , \label{bc_scalar_2}
\ee
with $\D(\D-3) = m^2$ and $\D$ chosen in such a way that the scalar
mode is normalizable. We solve Eq.~$(\ref{marginal})$ numerically
fixing $\alpha$ and $\beta$ and varying $L$ until we find a solution
with the correct asymptotic behavior $(\ref{bc_scalar_2})$. A
solution exists for $L\leq2\sqrt{3}$ and as we increase $L$ to lower
the temperature several marginally stable modes arise. Some examples
of marginal modes are shown in Figure \ref{fig:modes10} and
\ref{fig:modes20}.

The numerical results confirm the qualitative discussion at the
beginning of this Section. The instability temperature
$(\ref{Tc})$ provides a good estimate of the critical temperature at which
the first marginal mode actually appears, it overestimates
the numerical value by a factor of order one which decreases in the large $\alpha$ limit.
Marginal modes appear in a large region of the parameter space
and they provide a strong indication of the instability of the
AdS-RN black hole.

Finally integrating numerically Eq.~(\ref{scalar_pert}) with $\w \ne 0$ 
we found quasinormal modes with ${\rm Im}(\w) > 0$, which provide the main
indication for the existence of a dynamical instability. 
As expected the imaginary part of the frequency of these unstable modes 
increases as we increase $\a$ or consider values of $\b$ close to the BF bound.

\section{ Numerical solutions of the field equations}\label{sec:backreaction}

The existence of a perturbative instability for the AdS-RN black
hole in models containing a scalar field with a non-minimal coupling
to the gauge field suggests that the Lagrangian (\ref{lagr_gen})
should admit charged dilaton black holes with $T \le T_c$ and a
lower free energy than the AdS-RN black hole. In this Section we
show that these solutions exist by solving numerically the full
nonlinear equations of motion (\ref{max_scal_eq}). In Appendix
\ref{app:probe_limit} we discuss a simpler probe limit in which the
stress-energy tensors of the Maxwell and of the scalar field
decouple from the Einstein equations.

Analytic solutions for static, charged, planar black holes with
scalar hairs are difficult to find. Charged black holes with a
scalar hair in a similar model were found in flat spacetime
\cite{Monni:1995vu} and their generalization to AdS spacetime were
considered recently in \cite{Mignemi:2009ui}. The asymptotic
behavior of the scalar field is however not the one required for the
study of phase transitions in the dual theory. In
\cite{Stefanov:2007eq} numerical black hole solutions with a scalar
hair were found in flat space for models of dilaton gravity coupled
to the Born-Infeld action. Charged black hole solutions with scalar
hairs in AdS were only found resorting to numerical computations
\cite{Hartnoll:2008kx, Horowitz:2008bn} and we shall follow the same
approach here.

To set up the numerical procedure we first consider the behavior of the fields
near the AdS boundary
for $r \rightarrow \infty$ and near the horizon for $r \rightarrow
r_h$. The asymptotic behavior
near the AdS boundary of the scalar field is
\be \phi \sim \frac{{\cal O}_{-}}{r^{\D_-}}+\frac{{\cal
O}_{+}}{r^{\D_+}}\, , \label{bc_scalar} \ee
where \be \D_\pm=\frac{3\pm\sqrt{9+4m^2 L^2}}{2}\,.\ee 
In order to describe states of the dual field theory with a non
vanishing expectation value for the operator dual to the scalar
field, the asymptotic expansion $(\ref{bc_scalar})$ should contain
only normalizable modes \cite{Klebanov:1999tb}. For this reason when
$m^2 L^2 \ge -5/4$ we impose the boundary condition ${\cal O}_{-}
= 0$, corresponding to a vacuum expectation value for an operator of
dimension $\D_+$. When $- 9/4 < m^2 L^2 < -5/4$ two distinct choices
are possible \cite{Hertog:2004rz}: ${\cal O}_{-} = 0$,
corresponding to a vacuum expectation value for an operator of
dimension $\D_+$ and ${\cal O}_{+} = 0$, corresponding to a
vacuum expectation value for an operator of dimension $\D_-$.

The asymptotic behavior of the gauge field is
\be A_0 \sim \mu-\frac{\rho}{r}\, , \label{asymp_maxwell} \ee
where $\mu$ and $\rho$ specify the chemical potential and the charge
density of the dual theory \cite{Hartnoll:2009sz}. Finally the
asymptotic behavior of the metric functions is given by \be \chi
\sim \frac{ \D {\cal O}_{\D}^2}{4 L^2 } \, \frac{1}{r^{2\D}} \ , \ee
and \ba g &\sim& \frac{r^2}{L^2}+\frac{ \D {\cal O}_{\D}^2}{4 L^2}
\, \frac{1}{r^{2\D-2}} - \frac{2M}{r}\,, \hspace{1cm} {\rm if} \ 1 < 2 \D \le 2 \ , \nb \\
 g &\sim& \frac{r^2}{L^2} - \frac{2M}{r}\,, \hspace{3.4cm} {\rm if} \ \D > 1 \ , \label{asymp_mass}
 \ea
where $M$ is the black hole mass. Asymptotically the solution is then characterized
by four parameters: $\m$, $\r$, $M$ and ${\cal O}_\D$.

A power series expansion near the horizon shows that the solutions
of the equations $(\ref{eq:BR_scalar})$-$(\ref{eq:BR_einstein1})$
are completely specified by four parameters \cite{Hartnoll:2008kx}:
the horizon radius $r_h$, $A_0'(r_h)$, $\chi(r_h) \equiv \chi_h$ and
$\phi(r_h) \equiv \phi_h$. In term of these parameters the black
hole temperature is
 \be T=\frac{r_h}{16\pi
L^2}\left [ (12+2\phi_h^2)e^{-\chi_h/2} - L^2 A_0'(r_h)^2
e^{\chi_h/2} f(\phi_h) \right ]\,. \label{tempnum} \ee
In order to reduce the number of parameters in the numerical
analysis one can exploit the following three scaling symmetries of
the equations of motion  \cite{Hartnoll:2008vx} \ba r &\mapsto& k r
\ , \hspace{0.4cm} t \mapsto k t \ , \hspace{0.4cm} L \mapsto k L
\ , \nb \\
r &\mapsto& k r \ , \hspace{0.4cm} (t, x, y) \mapsto \frac{1}{k} (t,
x, y) \ , \hspace{0.4cm} g \mapsto k^2 g \ , \hspace{0.4cm} A_0
\mapsto k
A_0 \ , \nb \\
t &\mapsto& k t \ , \hspace{0.4cm} e^\chi \mapsto k^2 e^\chi \ ,
 \hspace{0.4cm} A_0 \mapsto \frac{1}{k} A_0 \ . \ea
We can use the first symmetry to set $L = 1$, the second to set $r_h
= 1$  (assuming that the horizon radius is different from zero) and
the third to set to zero the asymptotic value of $\chi$ at infinity.
In this way we are left with two parameters, $\phi_h$ and
$A'_0(r_h)$, that are constrained by the condition that either
${\cal O}_{+}$ or ${\cal O}_{-}$ vanish at infinity. It
follows that the numerical solutions of the field equations
$(\ref{eq:BR_scalar})$-$(\ref{eq:BR_einstein1})$ form a one-parameter
family. Varying the single free parameter, e.g. $\phi_h$, one
obtains solutions with different values of the temperature
($\ref{tempnum})$.

\begin{figure}[ht]
\begin{center}
\begin{tabular}{cc}
\epsfig{file=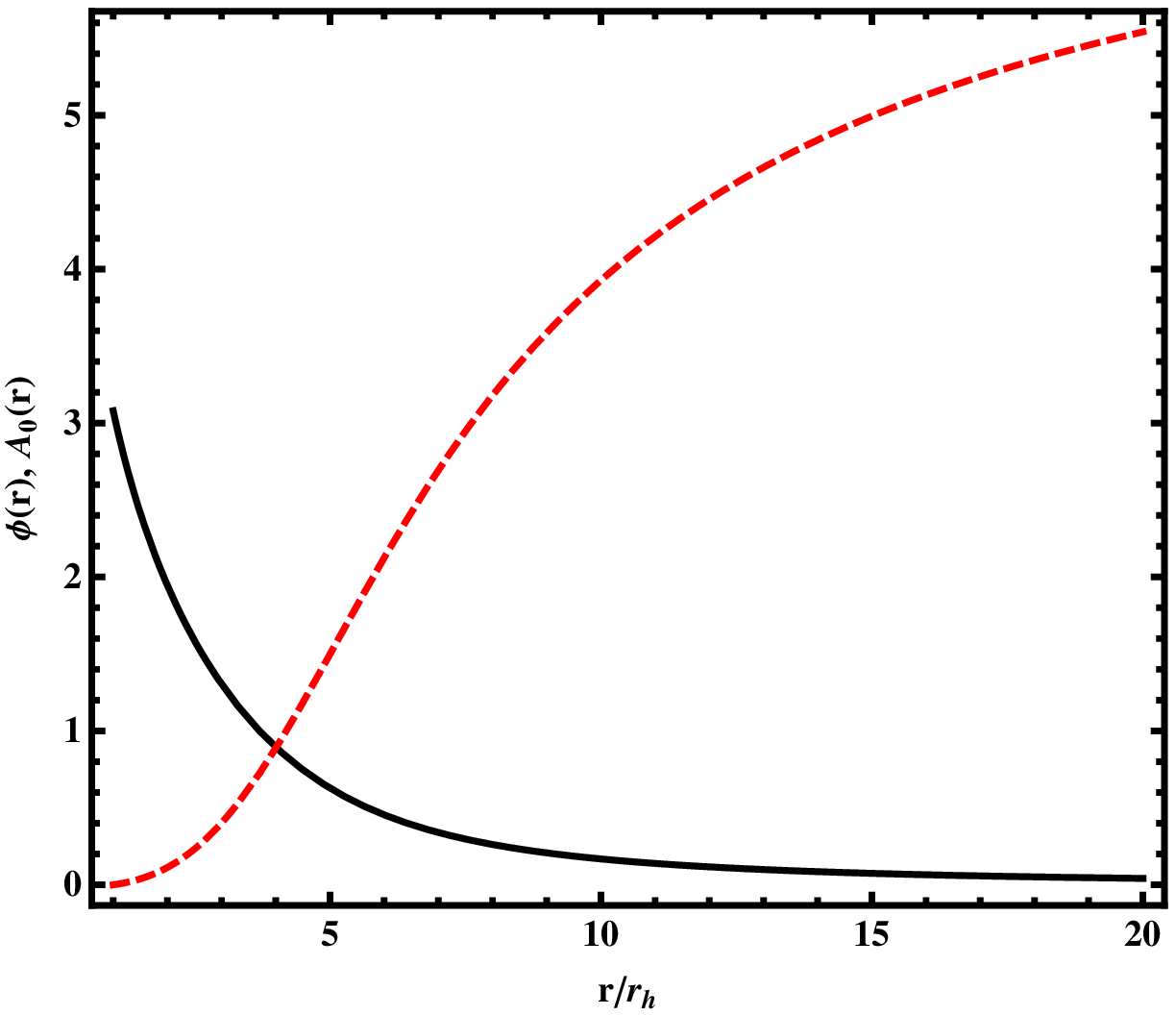,width=7cm,angle=0}&
\epsfig{file=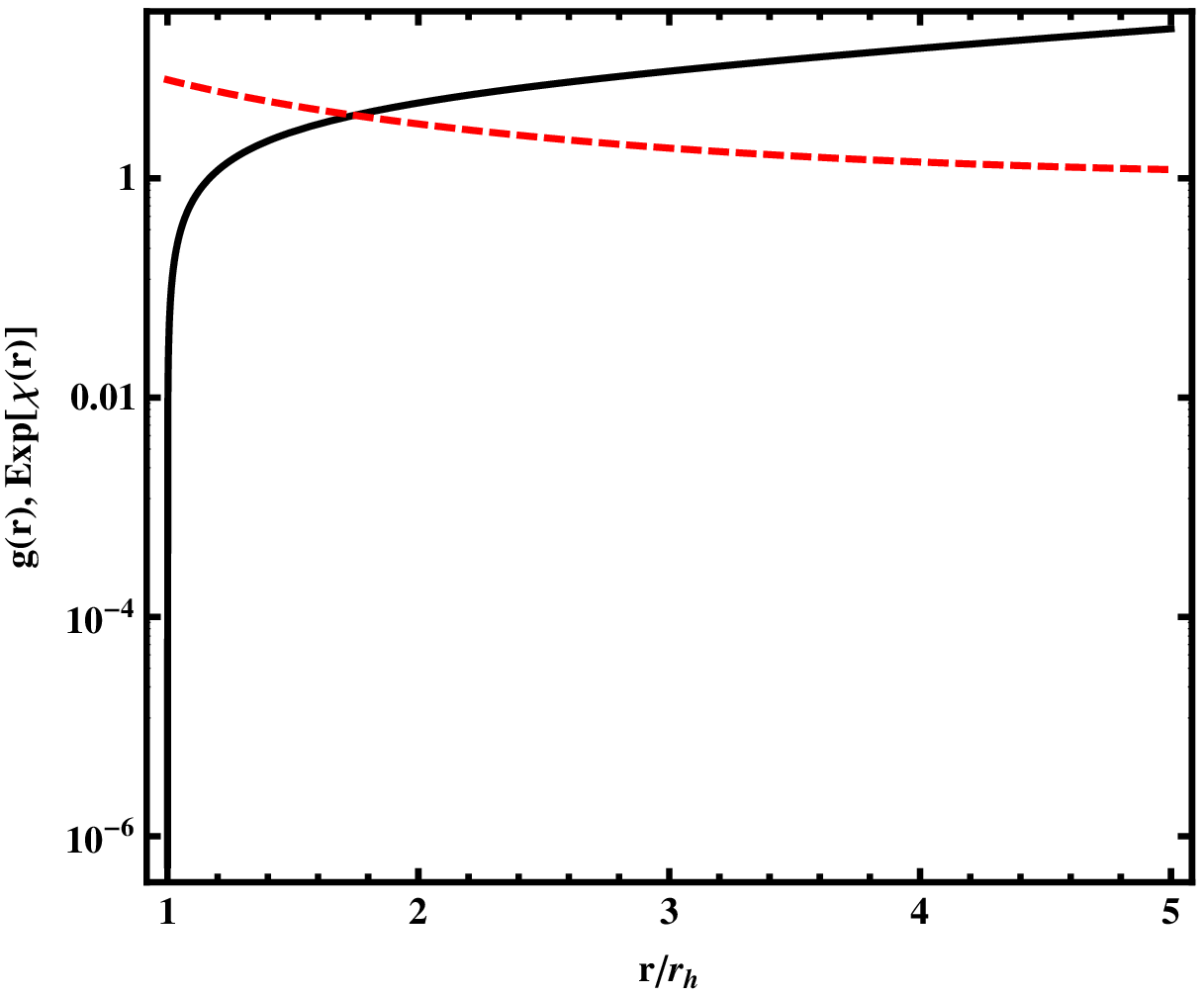,width=7cm,angle=0}
\end{tabular}
\caption{Examples of the field profiles of the numerical charged
black hole solution for a coupling function $f(\phi)= \cosh(2\phi)$
and a potential $L^2 V(\phi) = - 6 - \phi^2$. Left panel: the scalar
field $\phi$ (black) and the gauge potential $A_0$ (dashed red).
Right panel: metric function $g$ (black) and $e^{\chi}$ (dashed
red). $T/T_c \sim 0.2$.\label{fig:num_sol}}
\end{center}
\end{figure}

By numerical integration we found solutions of the field equations
$(\ref{eq:BR_scalar})$-$(\ref{eq:BR_einstein1})$ that describe
charged dilatonic black holes. We considered two types of coupling
functions \be f_{1}(\phi)= \cosh(a\phi) \ , \hspace{2cm}
f_{2}(\phi)= 1+\frac{\alpha}{2}\phi^2 \ , \ee and two types of
potentials \be V_{1}(\phi) = - \frac{6}{L^2} + \frac{\beta}{2
L^2}\phi^{2} \ , \hspace{2cm} V_{2}(\phi)= - \frac{6}{L^{2}}\cosh(b
\phi) \ . \label{modelpotentials} \ee In Figure \ref{fig:num_sol} we
display the field profiles of the numerical solution obtained for a
particular choice of $f$ and $V$. Other choices lead to
qualitatively similar results.

\begin{figure}[ht]
\begin{center}
\begin{tabular}{cc}
\epsfig{file=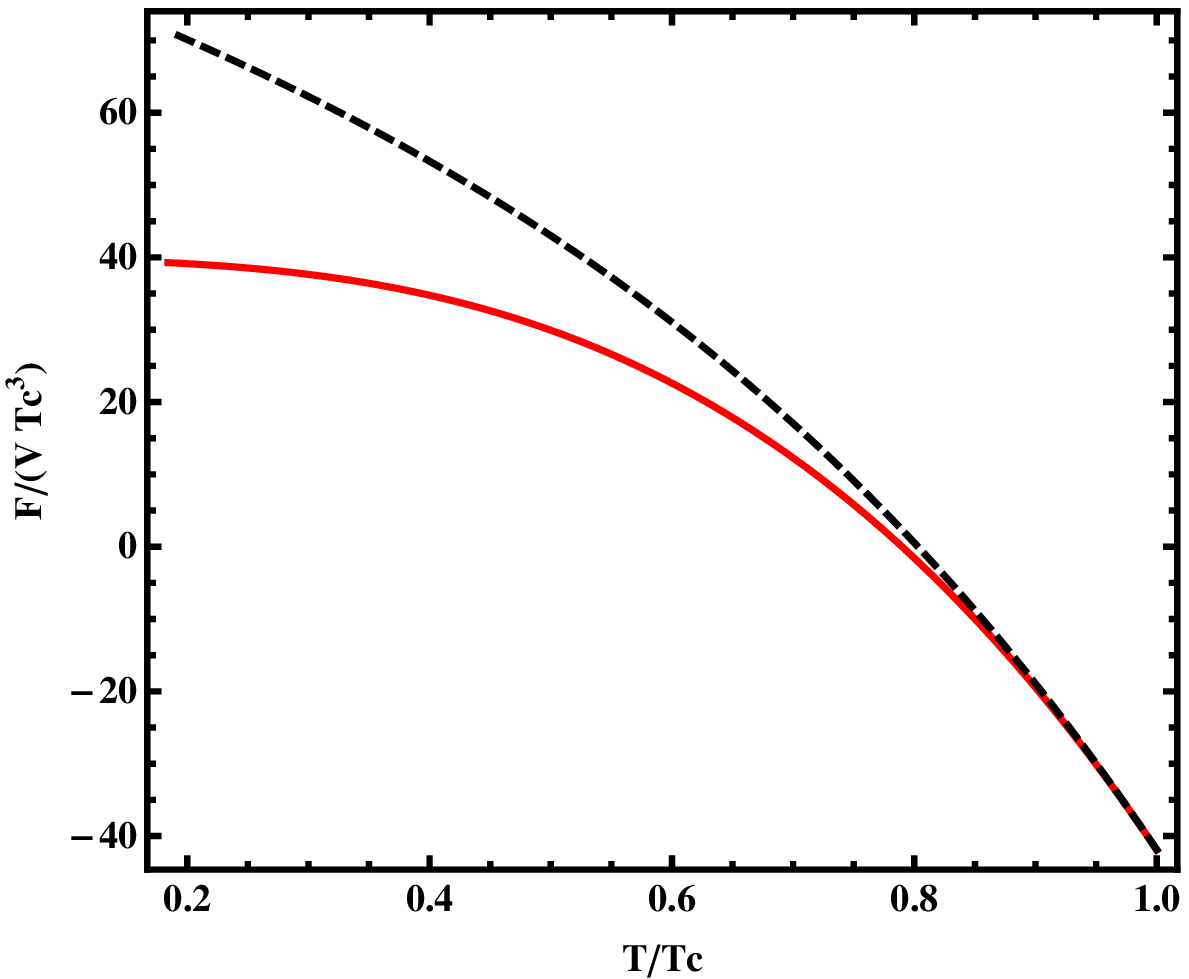,width=7cm,angle=0}&
\epsfig{file=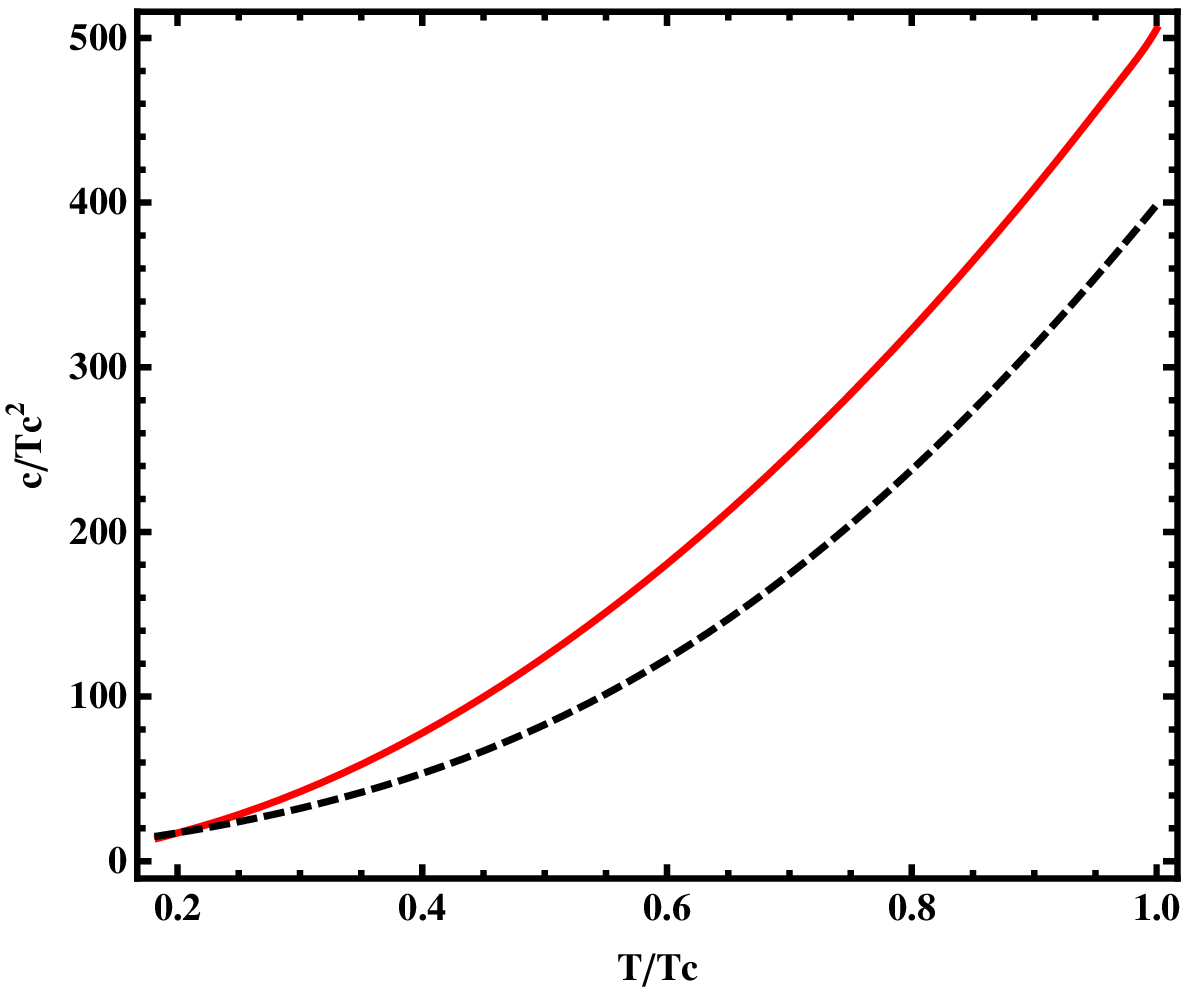,width=7cm,angle=0}
\end{tabular}
\caption{Left panel: Free energy of the hairy black hole (red line)
and of the AdS-RN (black dashed line).  Right panel: specific heat.  
The data shown are for the operator ${\cal O}_-$ 
and for $f(\phi)=\cosh(2\phi)$ and $V(\phi)=-6/L^2-\phi^2/L^2$. \label{fig:free_energy}}
\end{center}
\end{figure}

Below a critical temperature $T_c \sim \sqrt{\rho}$ we always find
charged dilaton black hole solutions.
Comparing their free energy
with the free energy of the AdS-RN one can verify that they
represent more stable states. The free energy $F=M-TS-\Phi Q$
depends on the four asymptotic parameters that characterize the
solution, namely $\m$, $\r$, $M$ and ${\cal O}_\D$. Setting $L=1$
the free energies of the AdS-RN black hole and of the dilaton black
hole read \cite{Hartnoll:2008kx}
\be F_{RN} =
\frac{V}{r_h}\left(-r_h^4+\frac{3\rho^2}{4}\right)\,,\hspace{2cm}
F_{CD} = V\left(-2M+\mu\rho\right)\, ,  \ee where $V$ is the volume
of the $(x,y)$ plane. In Fig.~\ref{fig:free_energy} we plot the free
energy and the specific heat $c=-T\partial^2_T F/V$ for an AdS-RN
black hole and a charged dilatonic black hole with the same mass and
charge. Below $T_c$ the dressed solution has a lower free energy and
represents a more stable state than the AdS-RN black hole. The free
energy is continuous at $T_c$ while the specific heat has a
discontinuity, so that the phase transition is second order.
For $T > T_c$ the dressed solution is not present anymore, precisely
as it happens for the instability induced by the minimal coupling to
a charged scalar field \cite{Hartnoll:2008vx,Ammon:2009fe}.
%
%
We mention here that for a given choice of $f$ and $V$ there are
usually several different black hole solutions with scalar hair and
the correct asymptotic behavior. We always choose the solution with
a monotonic scalar profile. The other solutions have scalar fields
with several nodes and presumably they also have a higher free
energy.

\section{The zero temperature limit}\label{sec:Tcon0}
Below the critical temperature the gravitational background
describes a charged dilaton black hole. In this section we discuss
the properties of the zero temperature ground state.
Extremal solutions are not easily found numerically. In order to
study their properties one can first look for an ansatz
for the leading behavior of the fields
near the horizon that contains at
least one free parameter. The equations of motion are then integrated
numerically and the free parameter varied until one finds the
correct asymptotic behavior at infinity. In this way a solution to
the equations of motion with the correct boundary conditions is
obtained \cite{Horowitz:2009ij}. The zero
temperature limit of hairy black holes in models with a minimally
coupled charged scalar field was studied in \cite{Horowitz:2009ij}.
In \cite{Goldstein:2009cv} a similar
analysis was carried out for dilaton black holes with
an exponential coupling function $f(\phi) = e^{a \phi}$.

We limit our attention to models with coupling function and
potential of the form \be f(\phi) = 2 f_0 \cosh a \phi \ ,
\hspace{1cm} V = - 2 W_0 \cosh b \phi \ . \ee A similar analysis
could also be performed for other coupling functions as for instance
$f(\phi) = 1 + \frac{\a}{2} \phi^2$. The action we consider is then
\be S = \int d^4 x \sqrt{-g} \left [ R - \frac{1}{2}(\p \phi)^2 -
\frac{f_0}{2} \cosh{a \phi} \, F^2 + 2 W_0 \cosh (b \phi) - 2
\Lambda \right ] \ , \ee where we included a negative cosmological
constant $\Lambda < 0$. The radius of the AdS vacuum solution is \be
\frac{3}{L^2} = W_0 - \Lambda \ , \ee and stability of the vacuum
requires that the parameter $b$ satisfies the Breitenlohner-Freedman
bound \be b^2 \le b_u^2 \ , \hspace{1cm} b^2_u \equiv
\frac{3}{8}\left[ 1 - \frac{\Lambda}{W_0} \right ] \ . \label{bf0t}
\ee In this section we use the following parametrization for the
metric \be ds^2 = - \lambda(r) dt^2 + \frac{dr^2}{\lambda(r)} +
H^2(r) (dx^2 + dy^2) \ . \ee The equation of motion of the gauge
field can be immediately integrated and gives \be A_0' = \frac{\r}{f
H^2} \ , \label{cond:gaugefield}\ee where $\r$ is the charge density
of the solution.
The remaining equations are \ba (\l H^2)'' &=& -2 H^2 \left(V+ 2 \Lambda \right)  \ , \nb \\
(H)'' &=& - \frac{H}{4}  (\phi')^2\ ,  \nb \\
(\l H^2 \phi')' &=&  H^2 \left [ \frac{dV}{d\phi} -
\frac{(A_0')^2}{2} \frac{d
f}{d \phi} \right ] \ , \nb \\
\l(H')^2 + \frac{\l'}{2}(H^2)' &=&  H^2 \left [ \frac{\l}{4}
(\phi')^2 -  \frac{f}{4} (A_0')^2 - \frac{V}{2} - \Lambda \right ] \
. \ea The equation of motion of the scalar field contains an
effective potential \be {\tilde V}(\phi) = V(\phi) + \frac{\r^2}{2
H^4 f(\phi)} \ , \ee and one can find $AdS_2 \times \mathbb{R}^2$
solutions with constant $H = H_0$ and $\phi = \phi_0$ provided that
\be \frac{d \tilde{V}}{d \phi}(\phi_0) = 0 \ , \hspace{1cm}
\frac{\r^2}{2 H_0^4 f(\phi_0)} =  - V(\phi_0) - 2 \Lambda \ . \ee
For the simple models discussed in this paper with potentials of the
form given in $(\ref{modelpotentials})$, this is possible only if
$m^2 > 0$ and for $ \r^2 a^2 > 4 f_0 m^2 H_0^4$. In this Section we
will focus on models with $m^2 \le 0$ and therefore we do not
consider this possible class of solutions.

\begin{figure}[ht]
\begin{center}
\begin{tabular}{cc}
\epsfig{file=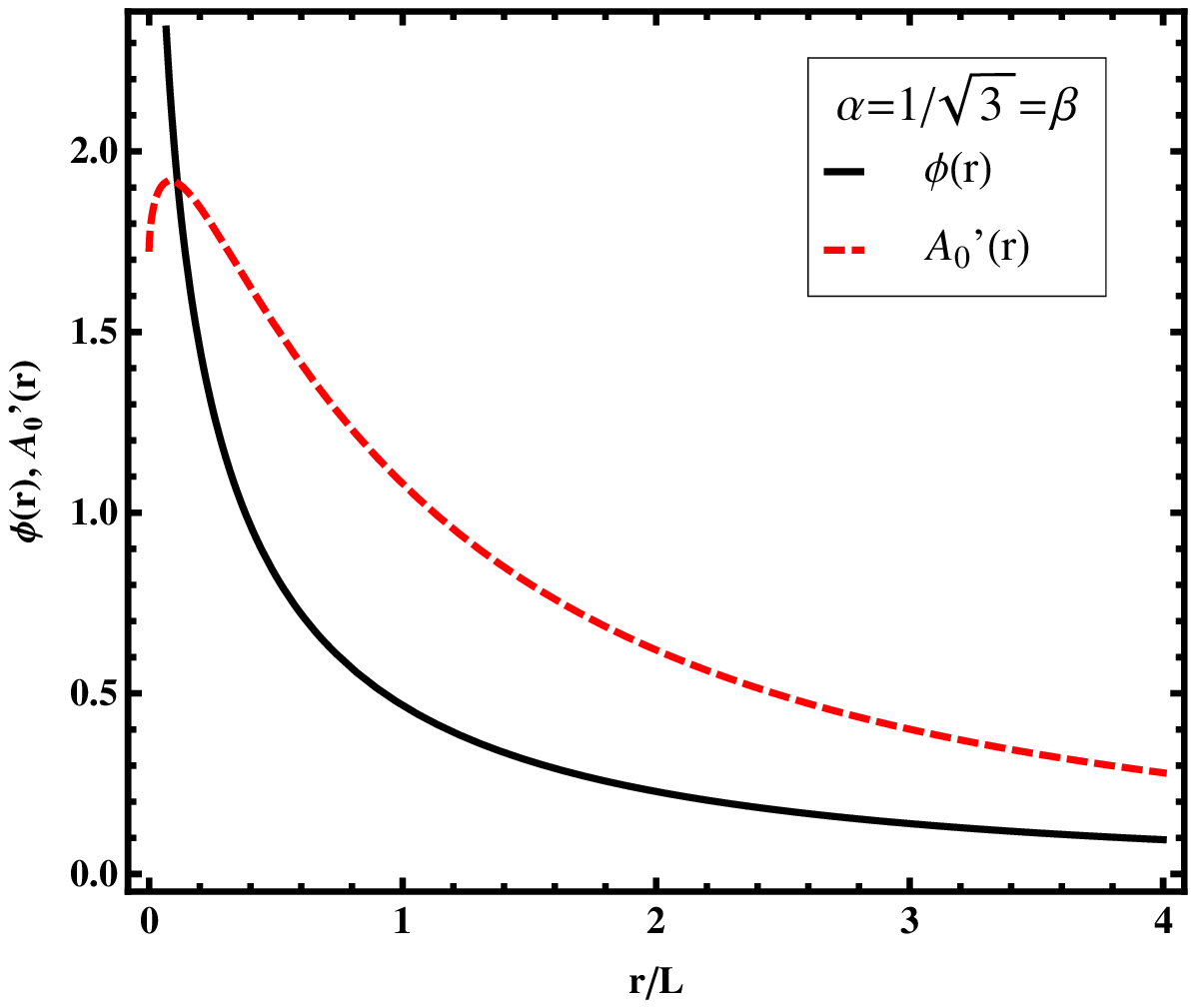,width=7cm,angle=0}&
\epsfig{file=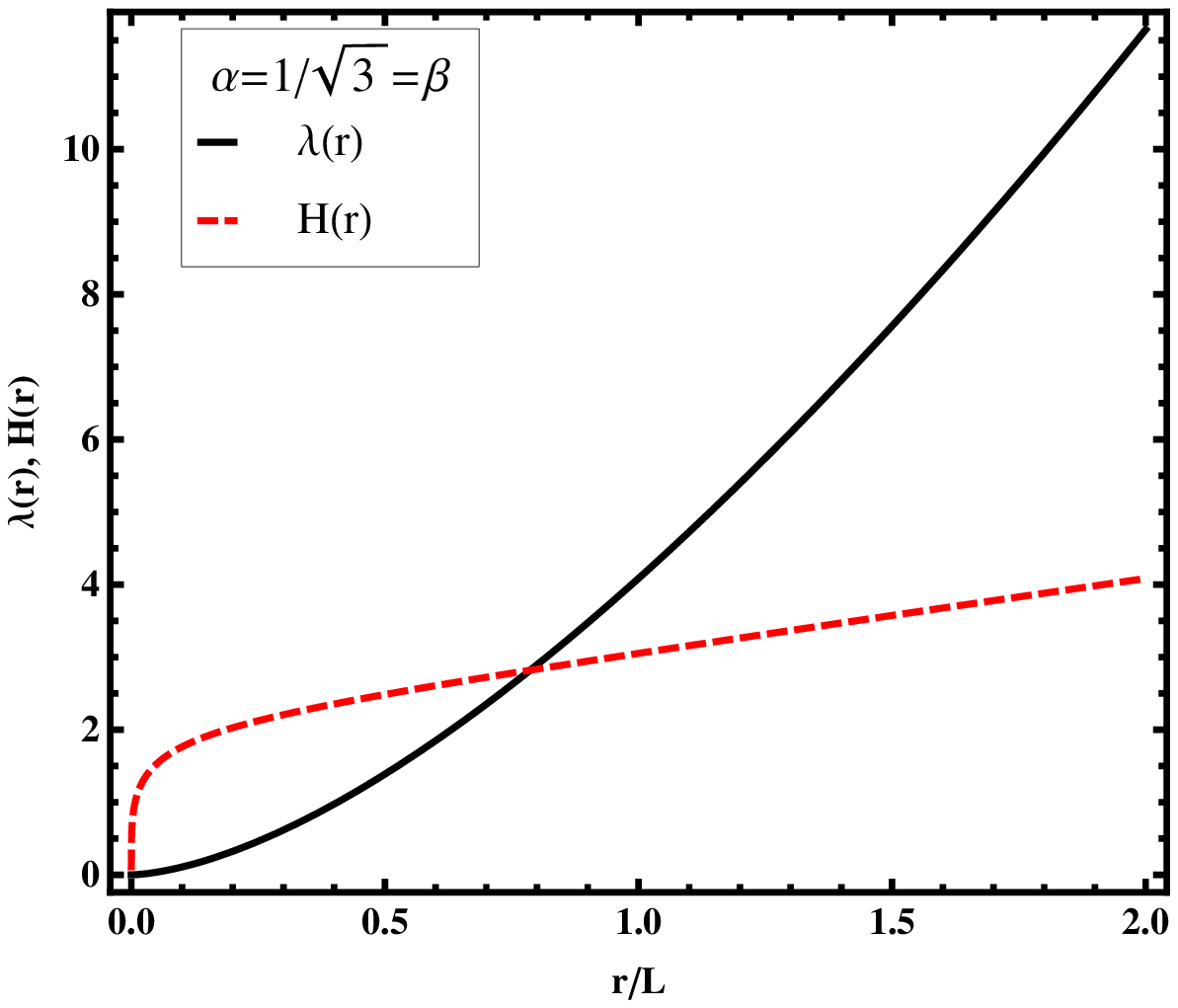,width=7cm,angle=0}
\end{tabular}
\caption{Background fields at $T=0$, Maxwell and scalar field (left)
and metric functions (right). \label{fig:T0_sol}}
\end{center}
\end{figure}

We will try instead the following scaling ansatz
\cite{Goldstein:2009cv} \be \l = \l_0 \, r^w ( 1 + p_1 \, r^\n ) \ ,
\hspace{1cm}  H = r^{h}(1 + p_2 \, r^\n) \ , \hspace{1cm} \phi =
\phi_0 - \xi \ln r + p_3 \, r^\n \ . \label{scalingansatz} \ee This
scaling ansatz provides an exact solution to the equations of motion
if the parameters are chosen in the following way \be \xi =
\frac{4(a+b)}{4+(a+b)^2}  \ , \hspace{1cm} w = 2 - b \xi \ ,
\hspace{1cm} h = \frac{(a+b)^2}{4+(a+b)^2} \ ,\ee \ba \l_0 &=&
\frac{2 W_0 e^{b \phi_0}}{(w+2h)(w+2h-1)} \ , \hspace{1cm}
\frac{\r^2}{f_0} e^{- a \phi_0} = \frac{2 W_0 e^{b \phi_0}}{w+2h}(2
- 2h-b \xi)  \ . \label{l0c0}  \ea Using the previous relations we
can express the parameter $\phi_0$ in terms of the charge density
and the other parameters of the model \be \frac{\r^2}{f_0} e^{- (a
+b) \phi_0} = 2W_0 \frac{2 - b(a+b)}{2+b(a+b)}  \ . \ee Finally the
other
parameters read \ba p_3 &=& \frac{2 p_2}{\xi} (2h - 1 + \n) \ ,  \\
p_1(w + 2h + \n)(w+2h-1+\n) &=& p_3 b(w+2h)(w+2h-1) - 2 p_2 \left
[\n^2 + \n (2w + 4h -1) \right ] \ , \nb \ea  and the exponent $\n$
is a real root of the following quartic equation \ba &&Q(\n) =
(\n+1)(\n+ a\xi)(\n^2+A\n+B) \ , \label{polly} \\ && A = 1+
\frac{2(a^2-b^2)}{4+(a+b)^2} \ , \hspace{0.4cm} B = -2A^2 + 2A
\frac{(a+b)^2(1+ab)}{4+(a+b)^2} \nb \ .  \ea If we choose the
greatest positive root the scaling ansatz describes the near-horizon
region of the extreme black hole. It could be interesting to study
the backgrounds corresponding to the other real roots of the
polynomial ($\ref{polly}$). In Figure \ref{fig:T0_sol} we show the
background fields obtained by integrating the near-horizon behavior
discussed above up to infinity and requiring suitable boundary
conditions.

The solution is extremal if $b \xi < 1$ which implies \be b < b_e \
, \hspace{1cm} b_e \equiv \frac{a}{3}\left [ 2
\sqrt{1+\frac{3}{a^2}} - 1 \right ] \ . \ee This condition together
with the BF bound $(\ref{bf0t})$ restricts the possible value of $b$
for a given value of $a$. When $b=0$ the previous solution reduces
to the solution found in \cite{Goldstein:2009cv} upon substituting
$W_0$ with $2(W_0 - \Lambda)$. Since the extremal solution depends
on the parameter $b$, it is not surprising that also the behavior of
the conductivity at low frequencies is modified. As we will show in
the next Section it vanishes as $\sigma(\w) \sim \w^{2 +
\frac{b\xi}{1-b\xi}}$.

The scaling ansatz in eq.~$(\ref{scalingansatz})$ can also capture
the near-horizon region of a near-extreme black hole
\cite{Goldstein:2009cv}. The near-extreme solution corresponds to a
negative value for the parameter $\n = - \h$  with \be \h = 1 +
\frac{\xi}{2}(a-b) \ , \ee which implies $p_2 = p_3 = 0$. The
parameters $\l_0$, $\r$ and $\f_0$ satisfy the relations already
given in (\ref{l0c0}). The solution depends then on a parameter $p_1
\equiv - m$ and the metric is \be ds^2 = \l_0 r^w \left ( 1 -
\frac{m}{r^{\h}} \right) dt^2 + \frac{dr^2}{\l_0 r^w \left ( 1 -
\frac{m}{r^{\h}} \right)} + r^{2-2h} (dx^2 + dy^2) \ . \label{nem}
\ee The horizon radius is given by $r_0^\h = m$. We will use this
metric in Section \ref{sec:holography} to clarify the low
temperature behavior of the optical conductivity.

We note that since the scalar field diverges logarithmically as it
approaches the extremal horizon, the same near-horizon analysis can
be applied to the case of exponential coupling functions of the form
$f(\phi) = f_0 e^{a \phi}$ and potentials of the form $V(\phi) = -
W_0 e^{b \phi}$. Black hole solutions in Einstein-Maxwell-dilaton
gravity with exponential coupling functions and Liouville-like
potentials are discussed in \cite{Cai:1996eg, Cai:1997ii,
Charmousis:2009xr}.

\section{Holographic properties of the new black hole solution}\label{sec:holography}

The instability of the AdS-RN black hole towards developing a scalar
hair signals a phase transition in the dual field theory defined on
the boundary of AdS. When the scalar field is charged with respect
to the $U(1)$ gauge field, in the dual theory there is a phase
transition to a superfluid state characterized by the spontaneous
breaking of a global $U(1)$ symmetry \cite{Hartnoll:2008vx}. In our
case the condensate is neutral and we have a phase transition
between the state dual to the AdS-RN black hole and the state dual
to the dilatonic black hole. As mentioned in the introduction, these
two states have markedly different properties, in particular from
the thermodynamical point of view.

In order to better illustrate the behavior of the new phase, we will display our results for
the four models listed in the following table
\begin{center}
\begin{tabular}{|l|l|l|l|l|} \hline
& $f(\phi)$ & $V(\phi)$ & $\alpha$ & $\beta$ \\  \hline  \ \
\emph{Model I}  \ \ &  \ \ $\cosh(\sqrt{3}\phi)$  \ \ & $ \ \
-\frac{6}{L^2}-\frac{\phi^2}{L^2}$  \ \ &
  $ \ \ 3 \ \ $ & $ \ \ -2 \ \ $ \\ \hline
 \ \  \emph{Model II}  \ \  & $ \ \ 1+\frac{3}{2}\phi^2$  \ \ &  \ \ $-\frac{6}{L^2}-\frac{\phi^2}{L^2}$  \ \ &
  \ \ $3$  \ \ &  \ \ $ -2$  \ \ \\ \hline
 \ \  \emph{Model III}  \ \  & $ \ \ \cosh(\sqrt{3}\phi)$  \ \ &
$ \ \ -\frac{6}{L^2}\cosh\left(\frac{\phi}{\sqrt{3}}\right)$  \ \ &
  $ \ \ 3 \ \ $ & $ \ \ -2 \ \ $ \\ \hline
 \ \  \emph{Model IV}  \ \  &  \ \ $\cosh(10\phi)$  \ \ &
 \ \ $-\frac{6}{L^2}-\frac{\phi^2}{L^2}$  \ \ &  \ \ $100$  \ \ &  \ \ $-2$ \ \  \\
\hline
\end{tabular}
\end{center}

The first three models correspond to different choices for the
coupling functions $f(\phi)$ and the potential $V(\phi)$ but they
all have the same values for the parameters $\alpha$ and $\beta$
that enter in the linear perturbation analysis of Section
\ref{sec:instability}. The last model illustrates the behavior of
the system for large values of the parameter $\alpha$. As discussed
in the previous Section, at low temperatures a simpler pattern
appears: when the potential has the form given in
$(\ref{modelpotentials})$ with $\b \le 0$ the properties of the
model depend mainly on $\a$ and are affected by the presence of the
potential only when this is an exponential function of the scalar
field.

\subsection{Phase transitions in the dual theory}

\begin{figure}[ht]
\begin{center}
\begin{tabular}{cc}
\epsfig{file=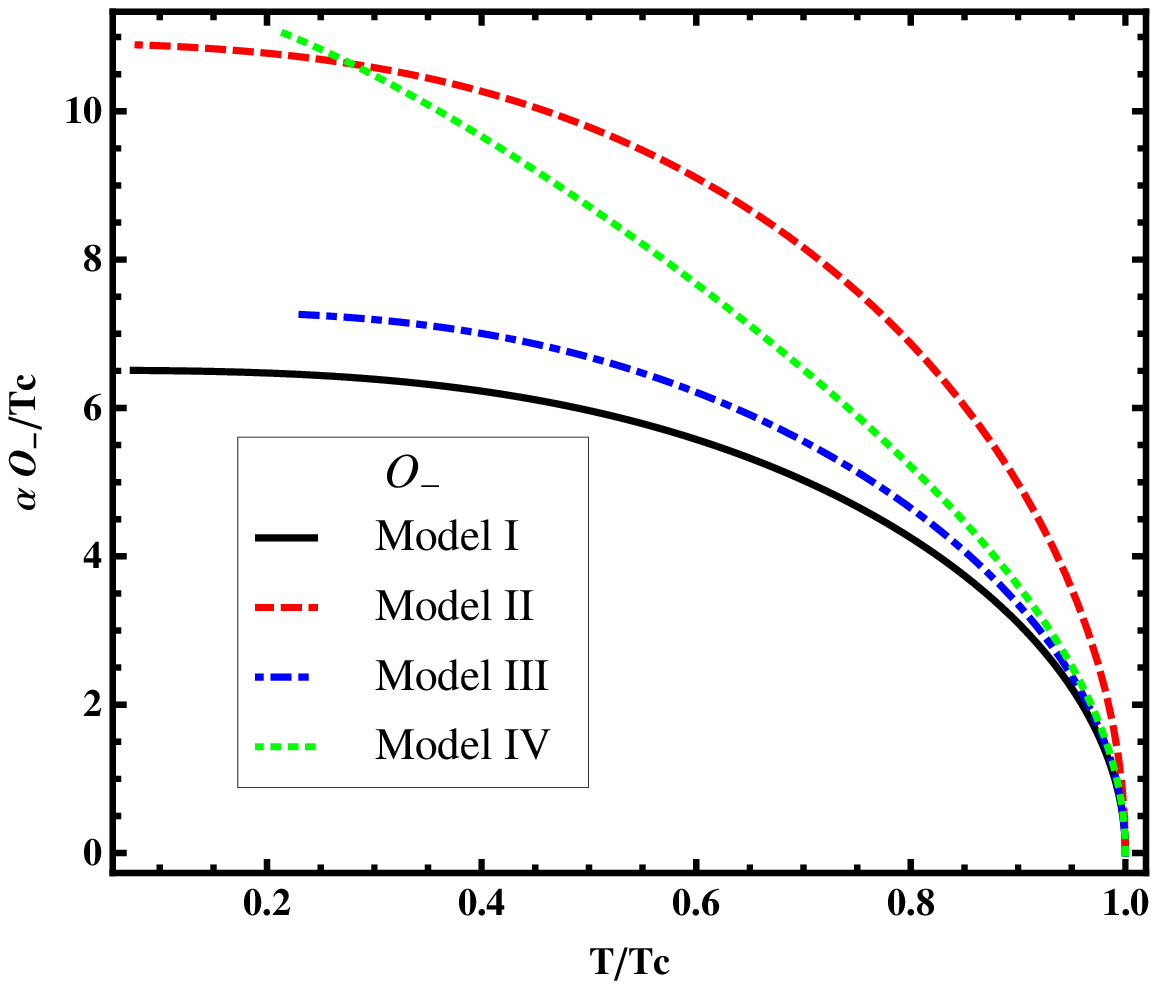,width=7cm,angle=0}&
\epsfig{file=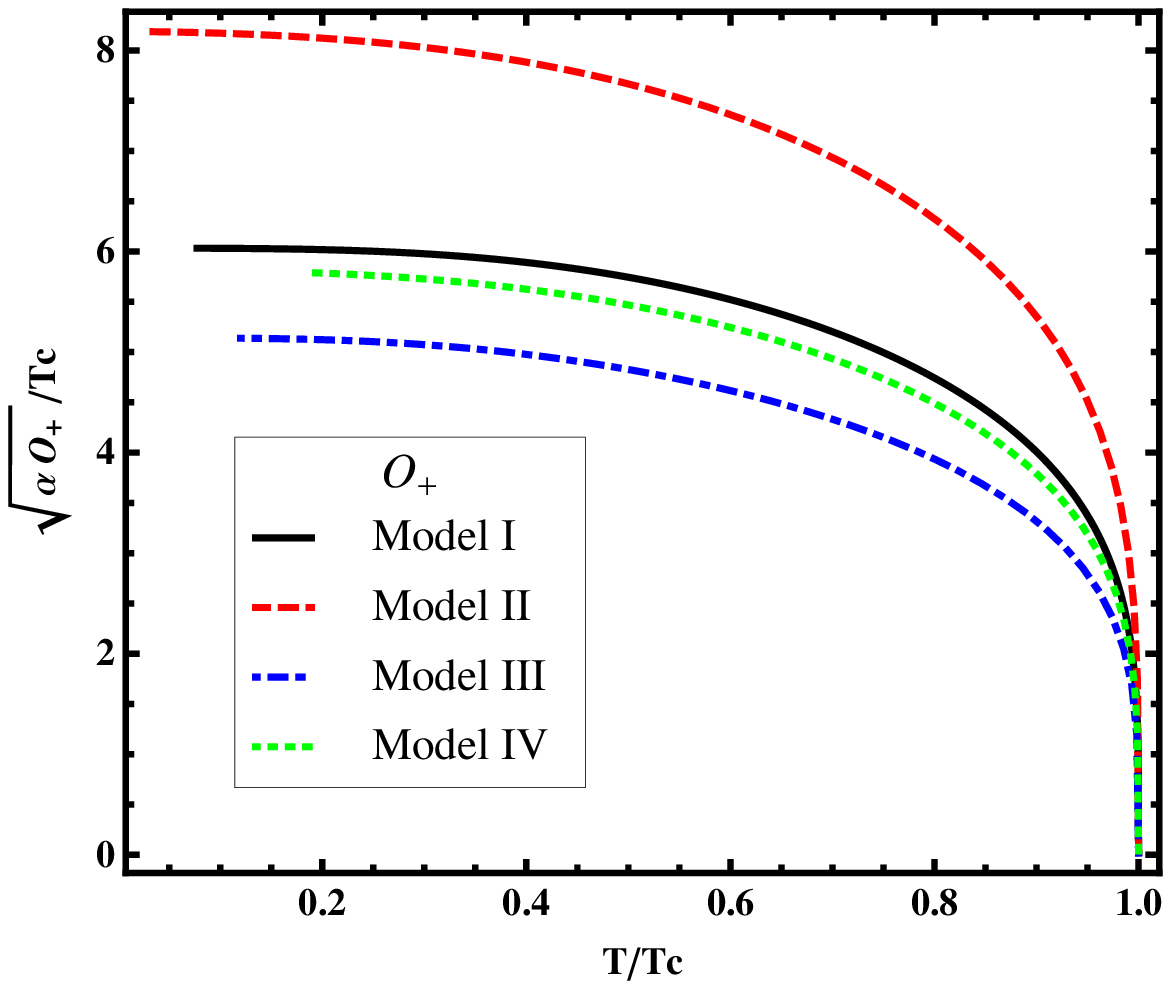,width=7cm,angle=0}
\end{tabular}
\caption{ Neutral condensate as a function of $T$ for the two
different boundary theories and for different models. Left: ${\cal
O}_{-} \sim(1-T/T_c)^{1/2}$ near the critical temperature. Right:
${\cal O}_{+} \sim(1-T/T_c)^{1/2}$ near the critical temperature.
\label{fig:condensate_backreaction}}
\end{center}
\end{figure}

Below $T_c$ the new black hole solution develops a scalar hair and
therefore the new phase of the dual theory is characterized by a
scalar condensate. The temperature dependence of the expectation
value of the neutral operator ${\cal O}_{+}$ or ${\cal O}_{-}$ can
be determined using eq.~(\ref{bc_scalar}) and the fact that the
black hole temperature depends only on $\phi_h$, the value of the
scalar field at the horizon. In Figure
\ref{fig:condensate_backreaction} we display the scalar condensate
at constant charge density and for different values of the
temperature for our four models.
The critical temperature is proportional to $\sqrt{\rho}$ where $\rho$ is the charge density.
Near the critical temperature the scalar condensate
behaves like $\sim(1-T/T_c)^{\gamma}$, with $\gamma=1/2$ for both
${\cal O}_{+}$ and  ${\cal O}_{-}$, which is the value of the exponent
predicted by mean field theory
\cite{Hartnoll:2008kx,Maeda:2009wv}.

\subsection{Electric conductivity in the dual theory}

\begin{figure}[ht]
\begin{center}
\begin{tabular}{cc}
\epsfig{file=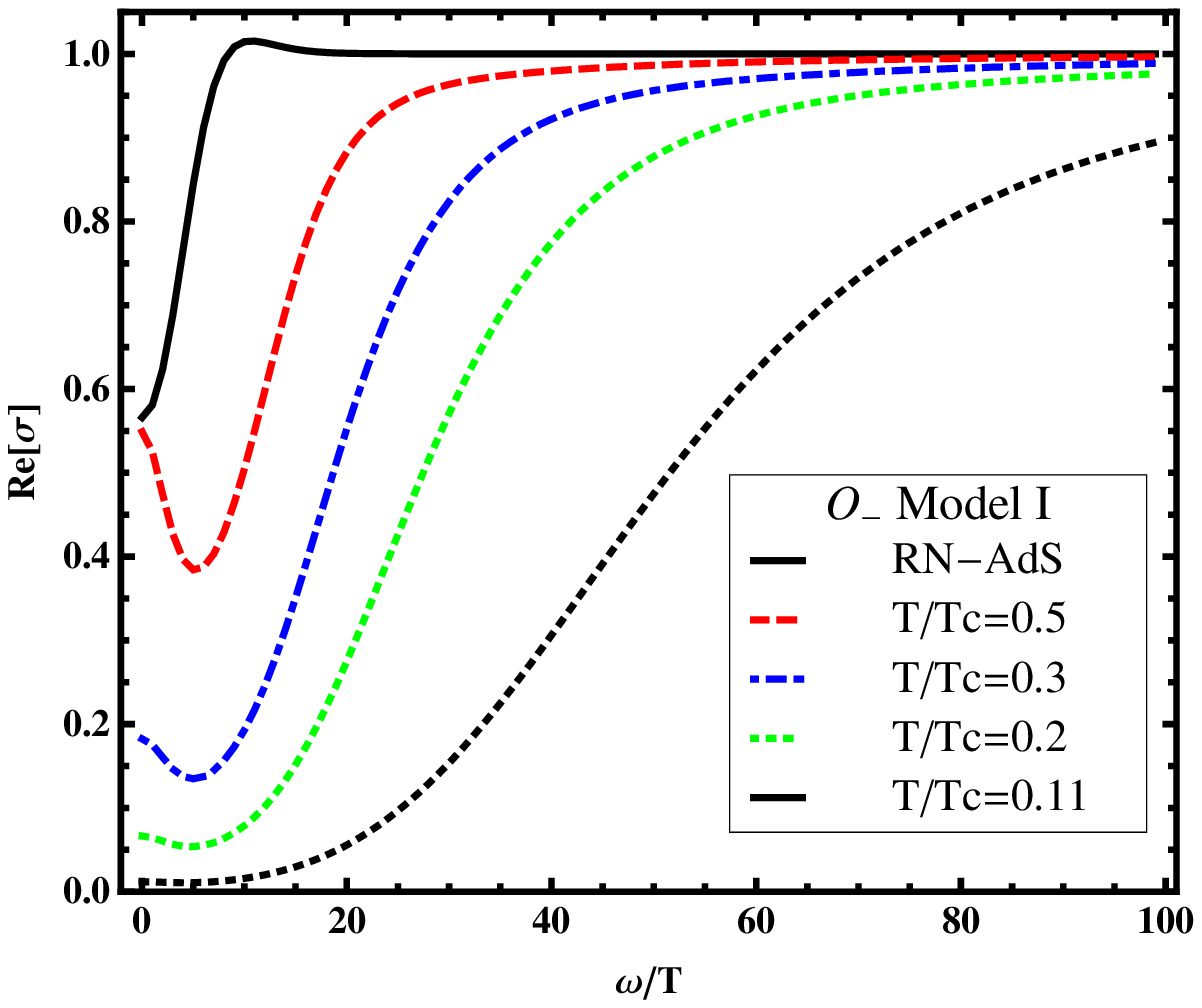,width=7cm,angle=0}&
\epsfig{file=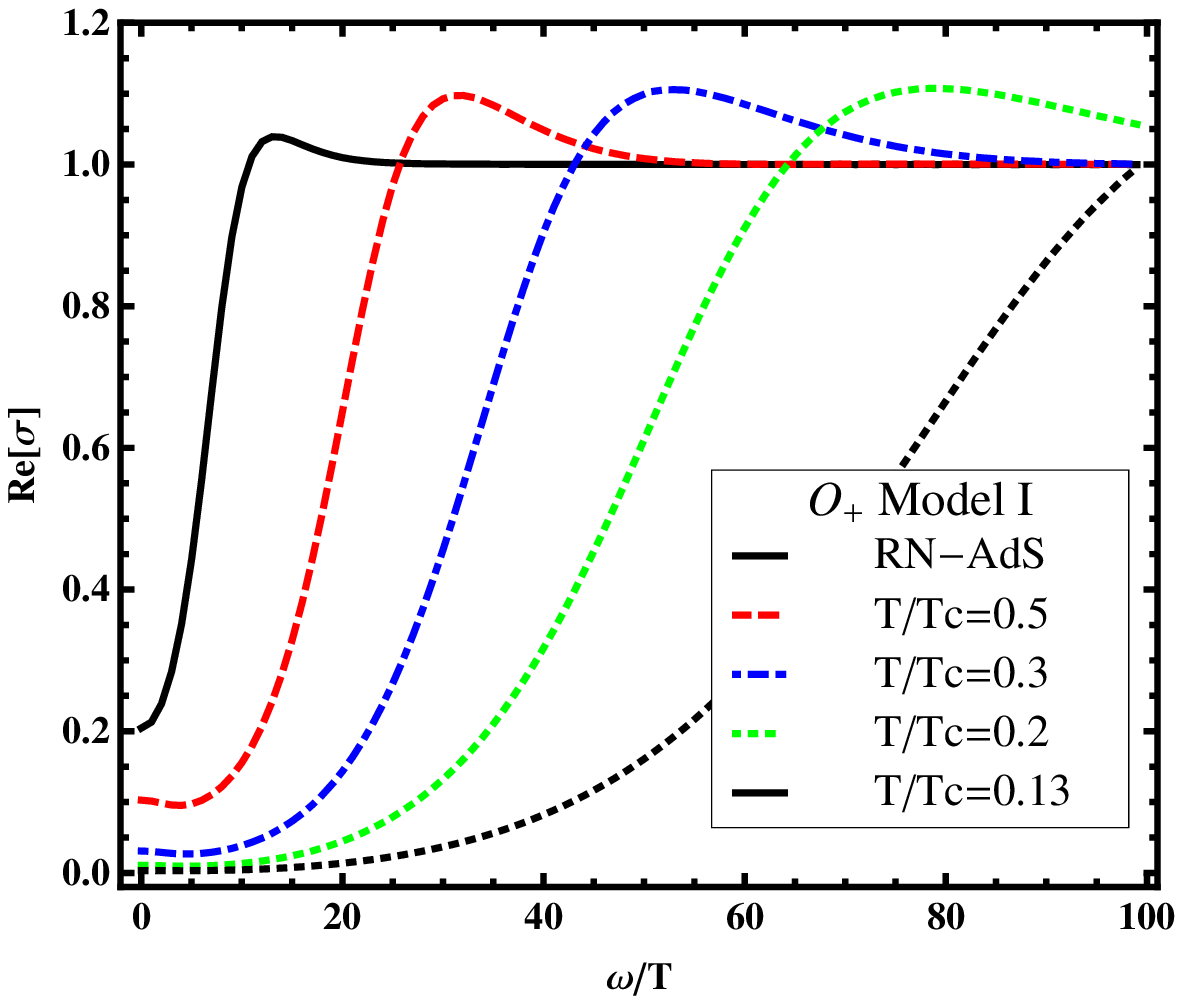,width=7cm,angle=0}
\end{tabular}
\caption{Real part of the conductivity as a function of the
frequency for different values of $T$ and for the operators ${\cal  O}_{-}$
(left) and ${\cal  O}_{+}$ (right). We show results for Model I.
\label{fig:conductivityBR1}}
\end{center}
\end{figure}

\begin{figure}[ht]
\begin{center}
\begin{tabular}{cc}
\epsfig{file=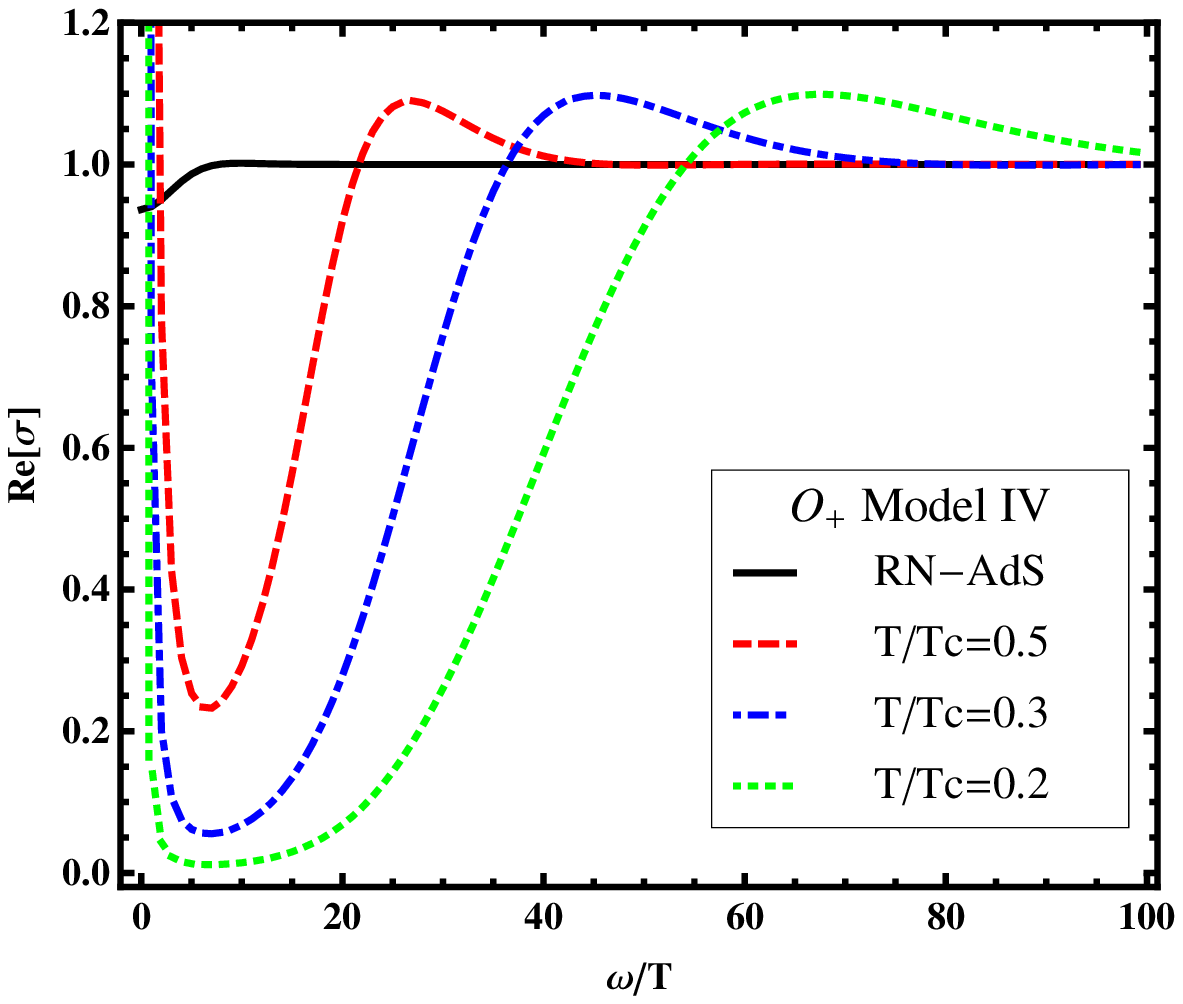,width=7cm,angle=0}
\epsfig{file=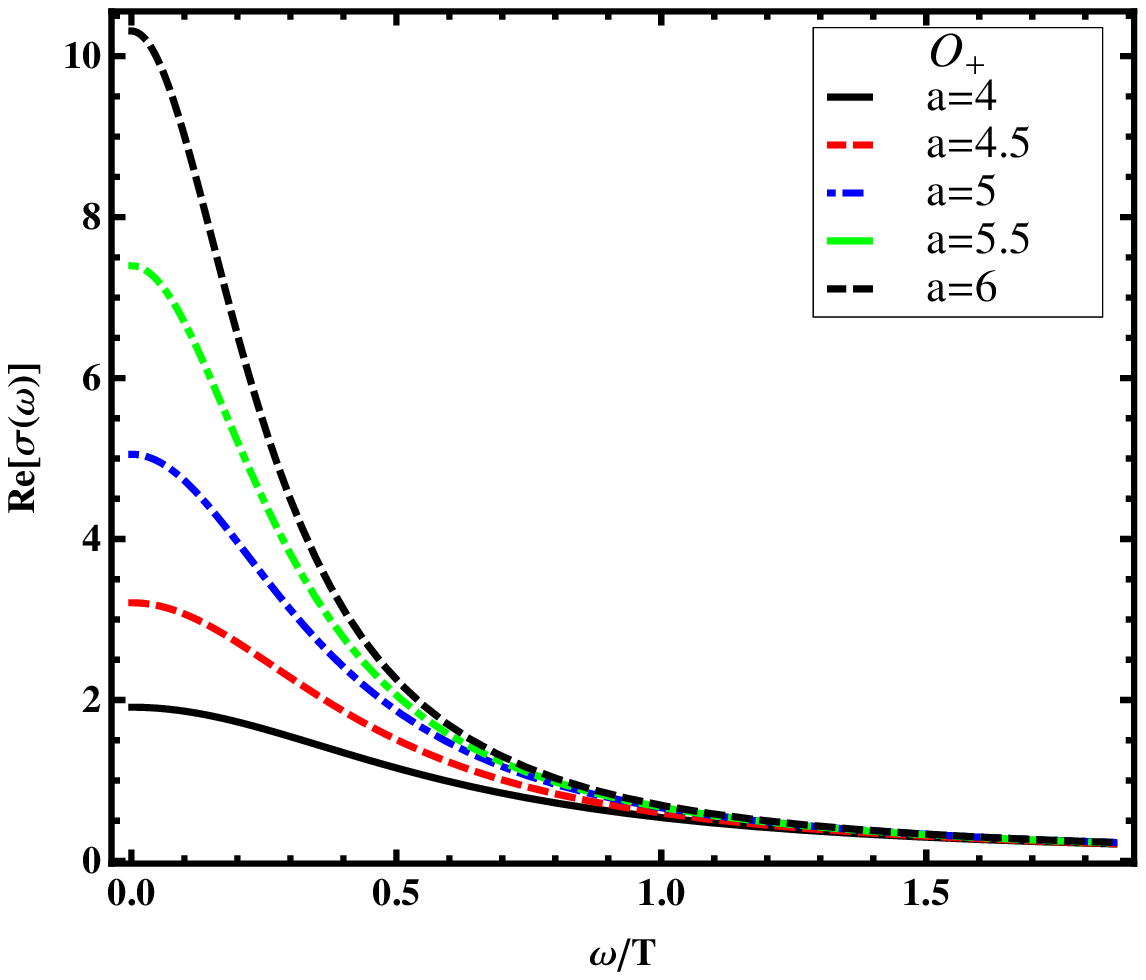,width=7cm,angle=0}
\end{tabular}
\caption{Left: Real part of the conductivity as a function of the
frequency for different values of $T$ and for the operator ${\cal  O}_{+}$.
We show results for Model IV. Right: Real part of the conductivity for
$\beta = -2$ and $f(\phi) = \cosh(a \phi)$ for increasing values of $a$.
\label{fig:conductivityBR1_modelIV}}
\end{center}
\end{figure}

\begin{figure}[ht]
\begin{center}
\begin{tabular}{cc}
\epsfig{file=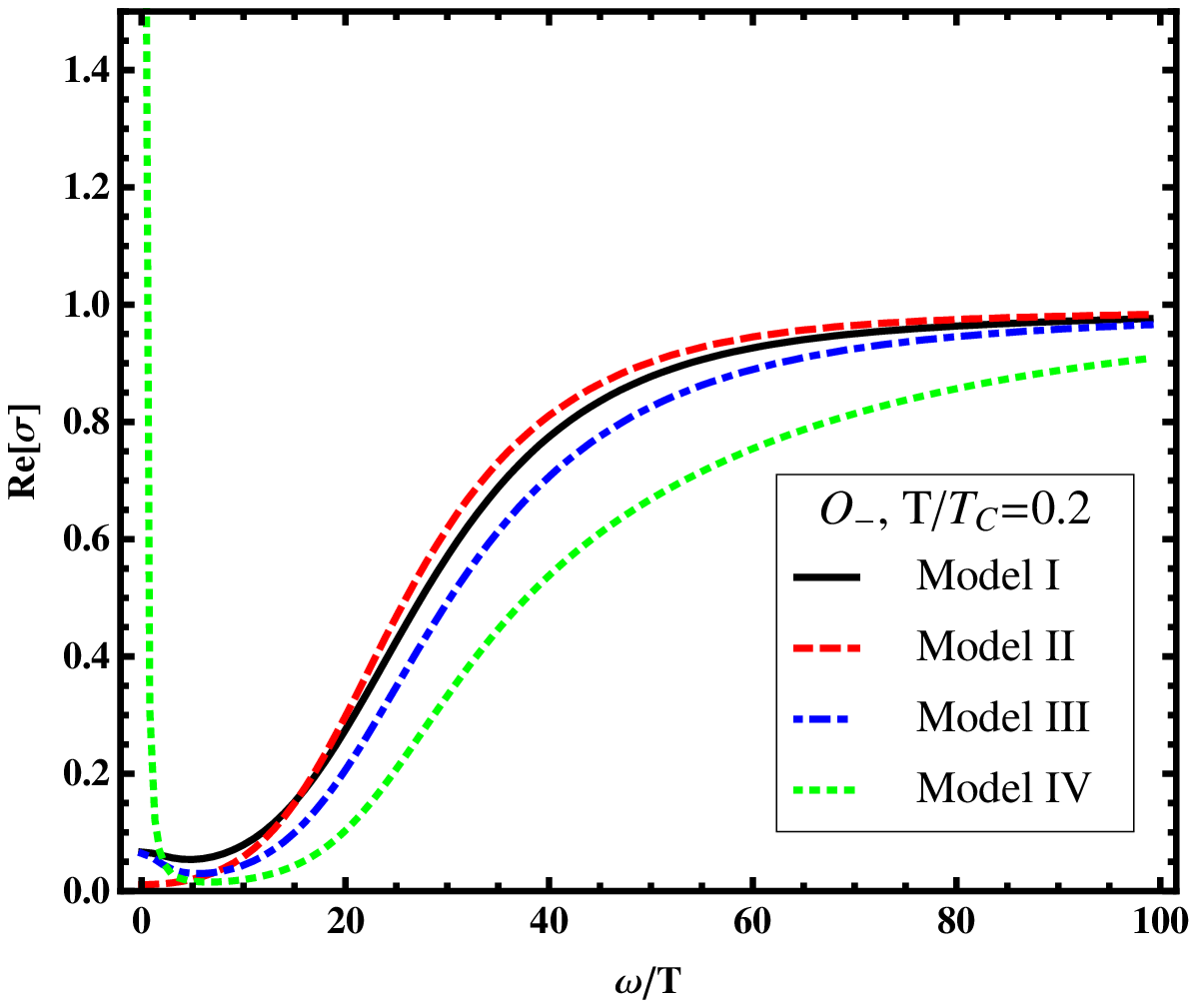,width=7cm,angle=0}&
\epsfig{file=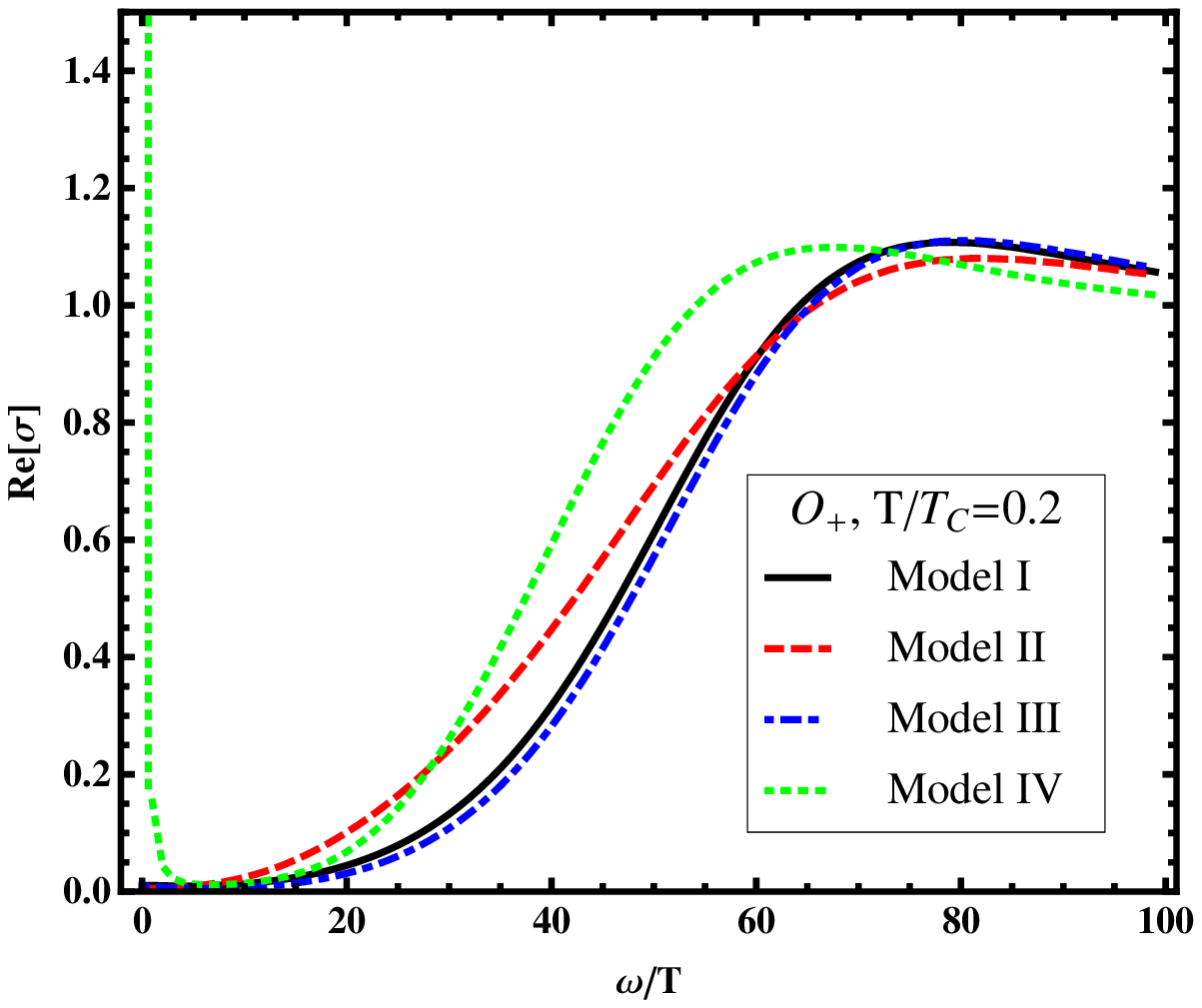,width=7cm,angle=0}
\end{tabular}
\caption{Real part of the conductivity as a function of the
frequency for different models at $T/T_c\sim0.2$ and for the operators
${\cal  O}_{-}$ (left) and ${\cal  O}_{+}$ (right).
\label{fig:conductivityBR2}}
\end{center}
\end{figure}

According to the AdS/CFT correspondence, transport phenomena in the
dual field theory are related to linear perturbations of the
equations of motion of the bulk fields. For instance  the electric,
thermal and thermoelectric conductivities can be derived from the
equations governing the fluctuations of the component $g_{tx}$ of
the metric and $A_x$ of the gauge field \cite{Hartnoll:2009sz}.
Perturbations of $g_{tx}$ and $A_x$ with zero spatial momentum and
harmonic time dependence decouple from all the other modes and one
is left with a system of just two equations
\ba A_x'' &+&
\left[\frac{g'}{g}-\frac{\chi'}{2}+\frac{1}{f(\phi)}\frac{df(\phi)}{d
\phi}\phi'\right]A_x'+\frac{ \omega^2}{g^2}e^\chi A_x = \frac{A_0'
e^\chi}{g}\left[\frac{2}{r}g_{tx}-g_{tx}'\right]\,,\label{eq:maxwell_pert_BR}\\
g_{tx}' &-& \frac{2}{r}g_{tx}+A_0'A_xf(\phi) =
0\,.\label{eq:einstein_pert_BR} \ea
Substituting the second equation in the first gives the following
equation for the fluctuations of the gauge field \be
A_x''+\left[\frac{g'}{g}-\frac{\chi'}{2}+\frac{1}{f(\phi)}\frac{df(\phi)}{d
\phi}\phi'\right]A_x'+
\left(\frac{\omega^2}{g^2}-\frac{{A_0'}^2f(\phi)}{g}\right) e^\chi
\, A_x=0\, . \label{eq:vp} \ee
We solve eq.~(\ref{eq:vp}) with purely ingoing boundary conditions
at the horizon. The electric, thermal and thermoelectric
conductivities are given by \cite{Hartnoll:2008vx} \be
\sigma=-i\frac{A_x^{(1)}}{\omega A_x^{(0)}}\,,\qquad
\sigma_{te}=\frac{1}{T}\left(\frac{i\rho}{\omega}-\mu\sigma\right)
\,,\qquad \sigma_{t}=\frac{iM}{4\omega T}\, ,
\label{definition_sigma} \ee where $A_x^{(0)}$ and $A_x^{(1)}$  are
fixed by the asymptotic behavior of the fluctuation at infinity \be
A_x\sim A_x^{(0)}+\frac{A_x^{(1)}}{r} \ .\label{asymp_pert} \ee In
the following we will consider only the electric conductivity.
Figures \ref{fig:conductivityBR1}, \ref{fig:conductivityBR1_modelIV} and \ref{fig:conductivityBR2} show
its frequency dependence obtained by numerical integration of
eq.~(\ref{eq:vp}).
%
The four models display similar qualitative features. In the
high-frequency limit the real part of the conductivity becomes constant, a property common to
every theory with an $AdS_4$ dual. Moreover, as a consequence of
translation invariance, the imaginary part of $\sigma$ has a simple
pole at $\omega = 0$ and therefore the DC conductivity contains a
delta function contribution at $\omega = 0$. The conductivity
displays interesting behavior at small frequencies. From the
plots we see that it develops a minimum and then it
tends to a non-zero value as $\w \to 0$, more
pronounced for large values of the non-minimal coupling $\alpha$, as
shown in Fig.~\ref{fig:conductivityBR1_modelIV}.
This non-zero value of
$\text{Re}[\sigma]$ at $\w=0$ may be seen as an analogue of the Drude peak
in the conductivity of ordinary metals.
As we lower the temperature
this additional contribution to the DC
conductivity decreases and at zero temperature $\s(\w)$ vanishes
at low frequencies as a power with an exponent fixed by the geometry
of the near-horizon region, as we will show below.

The frequency dependence of the conductivity is easier to understand
if one rewrites Eq.~(\ref{eq:vp}) as a Schr\"odinger equation and
expresses $\s$ in term of a reflection coefficient
\cite{Horowitz:2009ij}. In order to do so we introduce a new
coordinate $z$ defined by \be \frac{dr}{dz} = g e^{-\frac{\chi}{2}}
\ , \ee and rescale the gauge field setting $\sqrt{f} A = \Psi$. In
the new coordinates the horizon is at $z = - \infty$, the boundary
of AdS at $z = 0$ and the field $\Psi$ satisfies the following
equation
\be
\frac{d^2 \Psi}{d z^2} + (\omega^2 - V(z)) \Psi = 0 \ ,
\hspace{1cm} V(z) = g f (A_0')^2 + \frac{1}{\sqrt{f}} \frac{d^2
\sqrt{f}}{dz^2} \ . \label{schroedinger}
\ee
The potential in Eq.~(\ref{schroedinger}) has two contributions.
The second term is due to the non-minimal coupling, it is present also
in the probe limit and it is always negative in the near-horizon
region. The first term is always positive and corresponds to the
backreaction of the metric.

To find solutions of the original problem with ingoing boundary
conditions at the horizon it is convenient to extend the definition
of the potential $V(z)$ to positive values of $z$ by setting $V(z) =
0$ for $ z > 0$. One then solves the one-dimensional Schr\"odinger
equation with potential $V(z)$ for a particle incident on the
potential barrier from the right \cite{Horowitz:2009ij}. For $z
\geq 0$ the wave function is \be \Psi(z)=e^{-i\omega z}+{\cal
R}e^{i\omega z}\,,\qquad z\geq0\, , \ee 
where ${\cal R}$ is the reflection coefficient. Using Eq.~$(\ref{asymp_pert})$ and $(\ref{definition_sigma})$  and the
definition $\Psi = \sqrt{f} A$ one obtains
\be
\sigma(\omega)=-\frac{i}{\omega}\frac{A_x^{(1)}}{A_0^{(0)}}=\frac{1-{\cal
R}}{1+{\cal R}}-\frac{i}{2 \omega} \left [ \frac{1}{f}\frac{d f}{dz}
\right ]_{z=0} \, . \label{sigmaR} \ee

The second term in the equation above contributes only to the
imaginary part of the conductivity. When $f$
starts with a term linear in $\phi$ this term vanishes for $\D > 1$, is
constant for $\D = 1$ and diverges for $\D < 1$. When $f$ starts
with a term quadratic in $\phi$ and therefore satisfies the
conditions in Eq.~(\ref{expVandf}) it vanishes for $\D > 1/2$ and it
is constant for $\D = 1/2$. The real part of the conductivity is completely determined
by the reflection coefficient ${\cal R}$,
hence by the form of the potential $V$ in the Schr\"odinger equation.

The behavior of $V$ at large $r$ depends on the
asymptotic expansion of the scalar field $\phi \sim O_\D/r^\D$ and on
the coupling function $f(\phi)$. When $f$ starts with a term linear
in $\phi$, as it is the case for $f = e^{a \phi}$, the potential is
\be V(z) \sim \r^2 z^2 + \frac{a}{2}\D(\D-1)O_\D(-z)^{\D - 2} \ ,
\ee and therefore it vanishes for $\D > 2$, is constant for $\D = 2$
and diverges for $1/2 < \D < 2$. When $f$ starts with a term
quadratic in $\phi$, as it is the case for $f = \cosh (a \phi)$, the
potential is \be V(z) \sim \r^2 z^2 +
\frac{a^2}{2}\D(2\D-1)O^2_\D(-z)^{2(\D - 1)} \ , \ee and therefore
it vanishes for $\D > 1$, is constant for $\D = 1$ and diverges for
$1/2 < \D < 1$.

For non extremal black holes the potential tends to zero exponentially
as the coordinate $z$ tends to the horizon, $V(z) \sim V_h e^{4 \pi T z}$ where $T$
is the temperature of the black hole.
While  the potential of a model
with  a charged scalar minimally coupled to the gauge field
is always positive, in our case the
sign of the potential is not definite.
Although it is always positive near the boundary, it can
become negative and reach a minimum before vanishing at
the horizon with a negative exponential tail.
Whether the
potential is positive or negative depends on the relative magnitude
of the two terms in the Schr\"odinger potential, the
one due to the non-minimal coupling and the one due to the backreaction.
As we lower the temperature the effect of the backreaction becomes
increasingly important. In fact at
$T=0$ the potential is always positive near the extremal horizon
and vanishes like $C/z^2$, with a positive constant $C$.

The near-horizon behavior of the potential at $T=0$ can be
determined using the analytic form of the extremal solution found in
the previous Section. With the form of the metric used in that
Section the new coordinate $z$ is given by \be \frac{dr}{dz} = \l \
, \ee and the rescaled gauge field $\Psi = \sqrt{f} A $ satisfies
the following equation \be \frac{d^2 \Psi}{d z^2} + (\omega^2 -
V(z)) \Psi = 0 \ , \hspace{1cm} V(z) = \l f (A_0')^2 +
\frac{1}{\sqrt{f}} \frac{d^2 \sqrt{f}}{dz^2} \ . \label{eqs2} \ee
Substituting the explicit form of the solution one obtains \be V(z)
= \frac{C}{z^2} \ , \hspace{1cm}  C = \frac{(2 b \xi - 3)^2}{4(1-b
\xi)^2} - \frac{1}{4} \ . \ee Following the approach of
\cite{Gubser:2008wz} one can show that the optical conductivity
vanishes at small frequencies and obtain an analytic expression for
the leading power in $\w$ by matching the conserved probability
current of the Schr\"odinger equation near the boundary at infinity
and near the horizon. The result is
 \be \s(\w) \sim
\w^{2 + \frac{b \xi}{1 - b \xi}}  \ , \ee and this behavior is
observed in our numerical results for the conductivity in the low
temperature and low frequency limits.

\begin{figure}[ht]
\begin{center}
\begin{tabular}{cc}
\epsfig{file=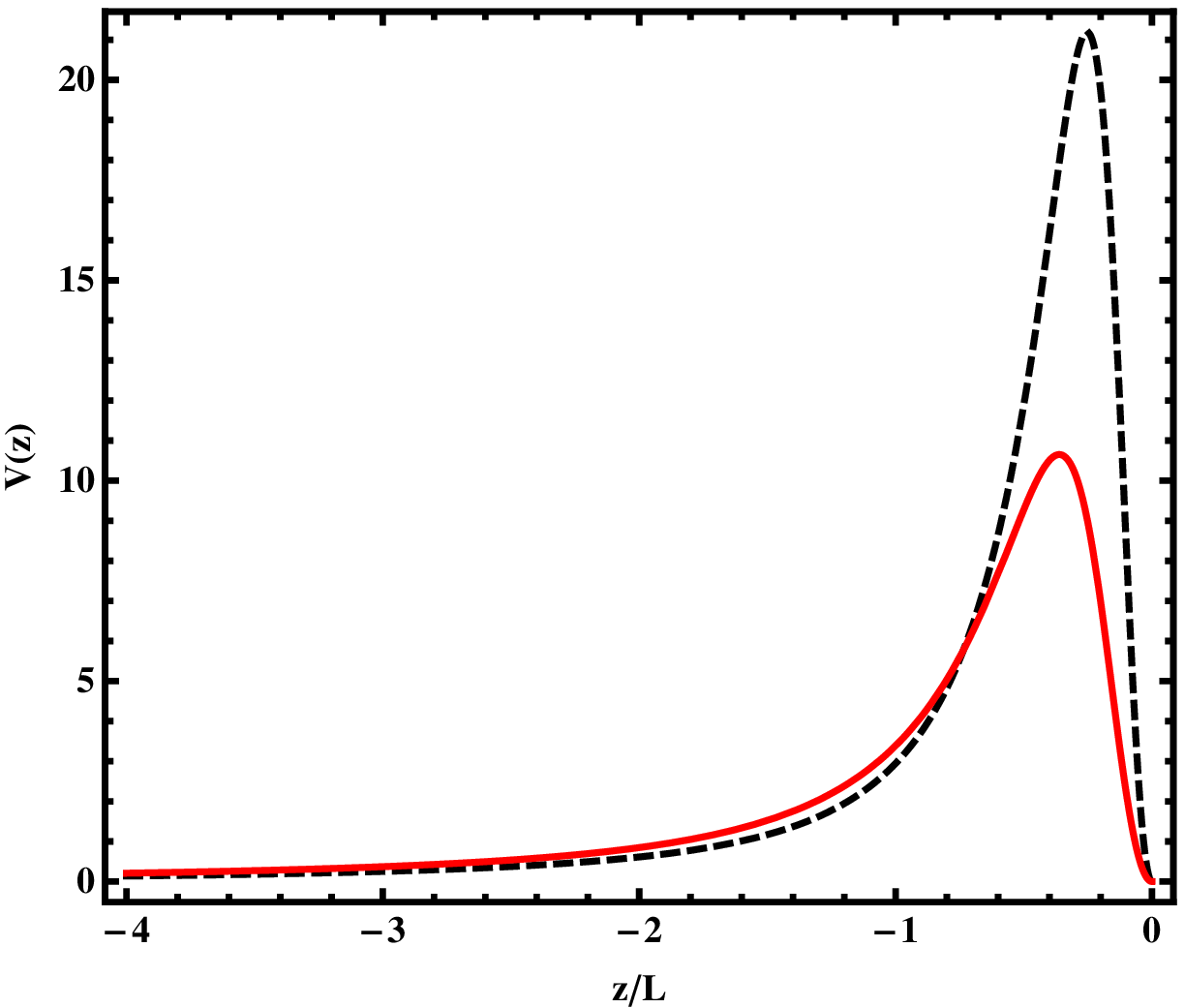,width=7cm,angle=0}&
\epsfig{file=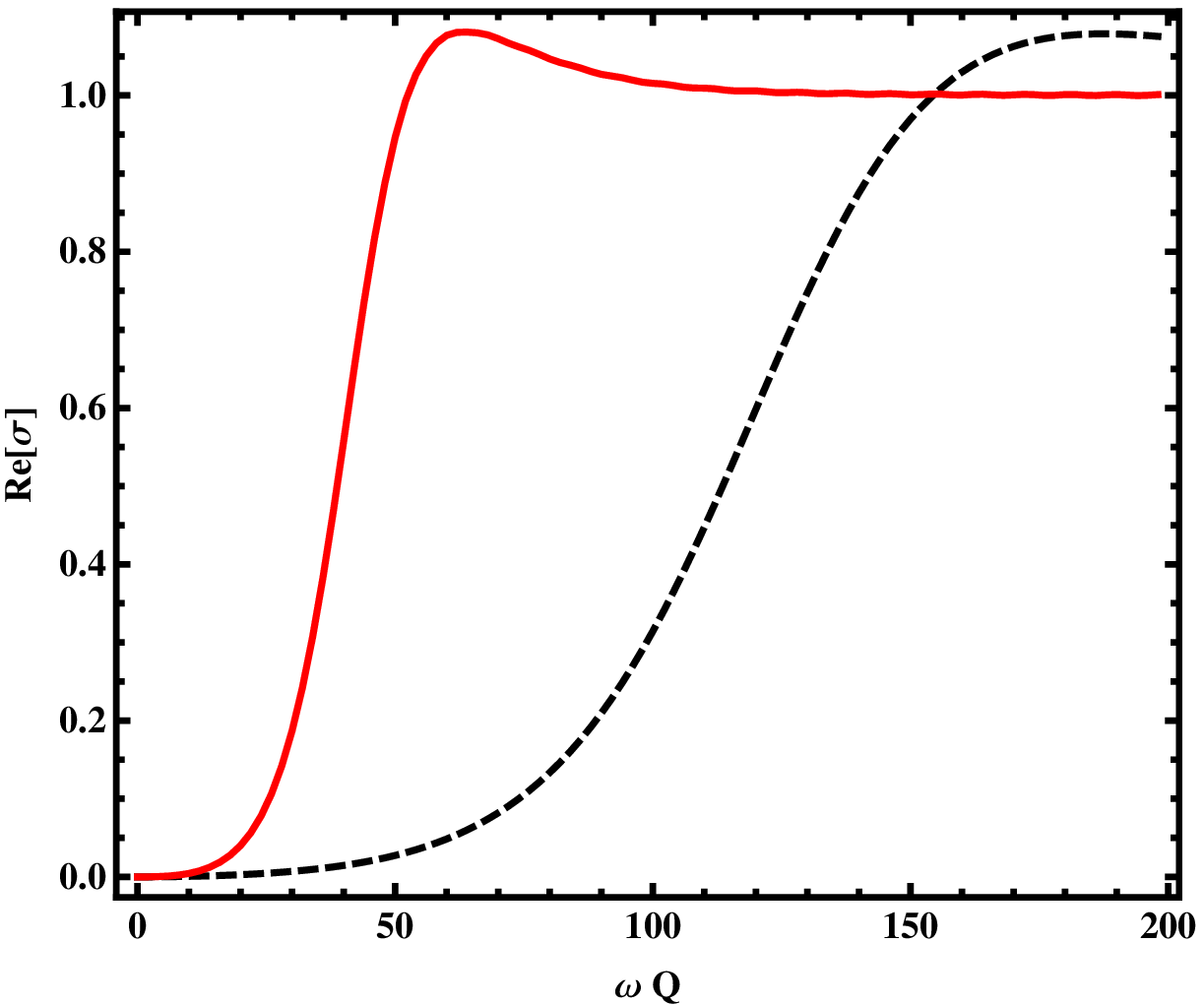,width=7cm,angle=0}
\end{tabular}
\caption{Schr\"odinger potential (left) and electrical conductivity
(right) at $T=0$. Black dashed lines correspond to $b=0$ and $a=1$,
whereas red straight lines correspond to $b=a=1/\sqrt{3}$.
\label{fig:T0_cond}}
\end{center}
\end{figure}

The Schr\"odinger potential and the related conductivity are shown in Figs.~\ref{fig:T0_cond}.
As mentioned before, also in this case there are several solutions to the field equations.
The Schr\"odinger potential and the conductivity for
these additional solutions can have several maxima and minima in the intermediate region
between the horizon and infinity. These features are absent if the coupling function is
exponential, $f\sim e^{a\phi}$.

\begin{figure}[ht]
\begin{center}
\begin{tabular}{cc}
\epsfig{file=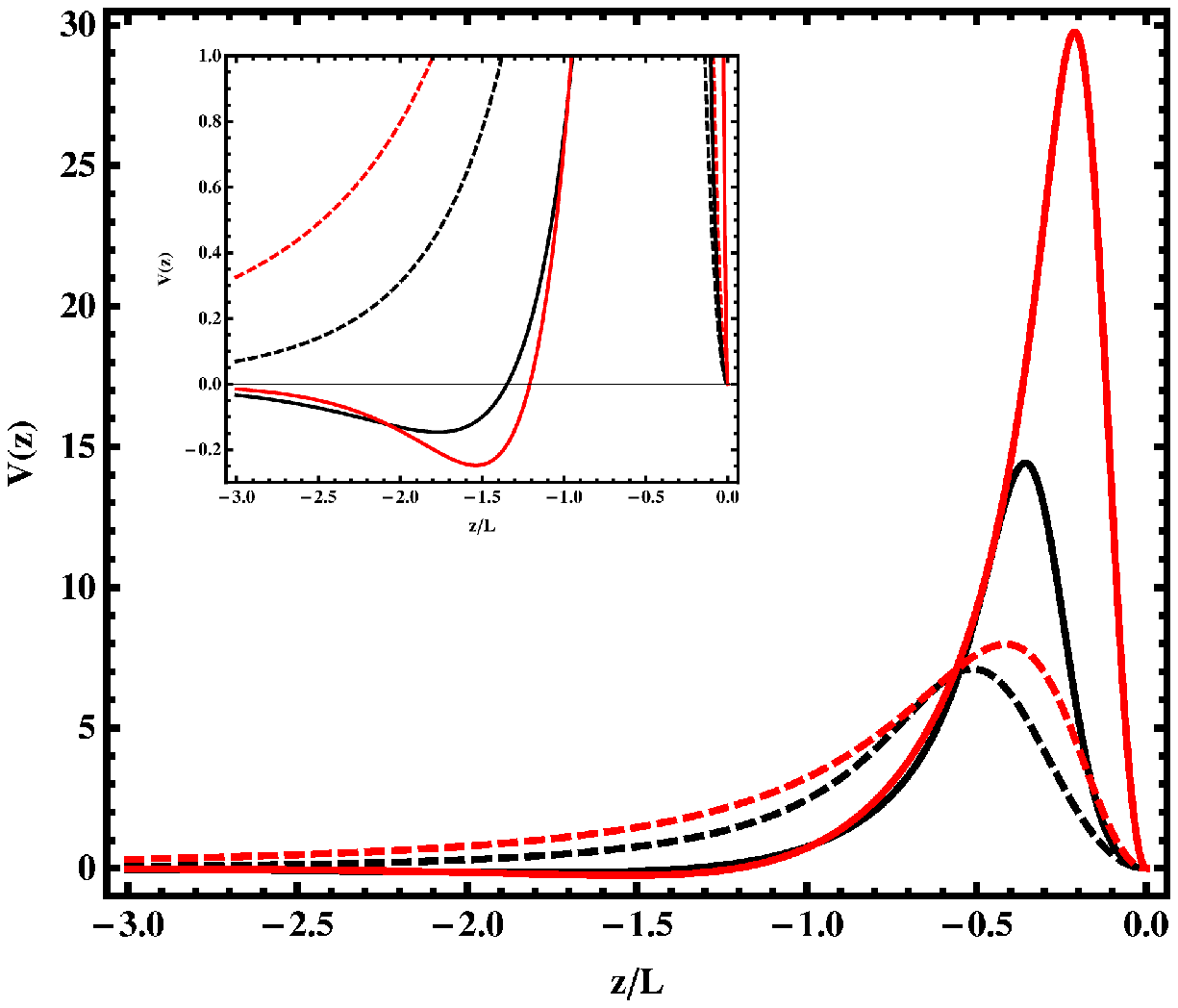,width=7cm,angle=0}&
\epsfig{file=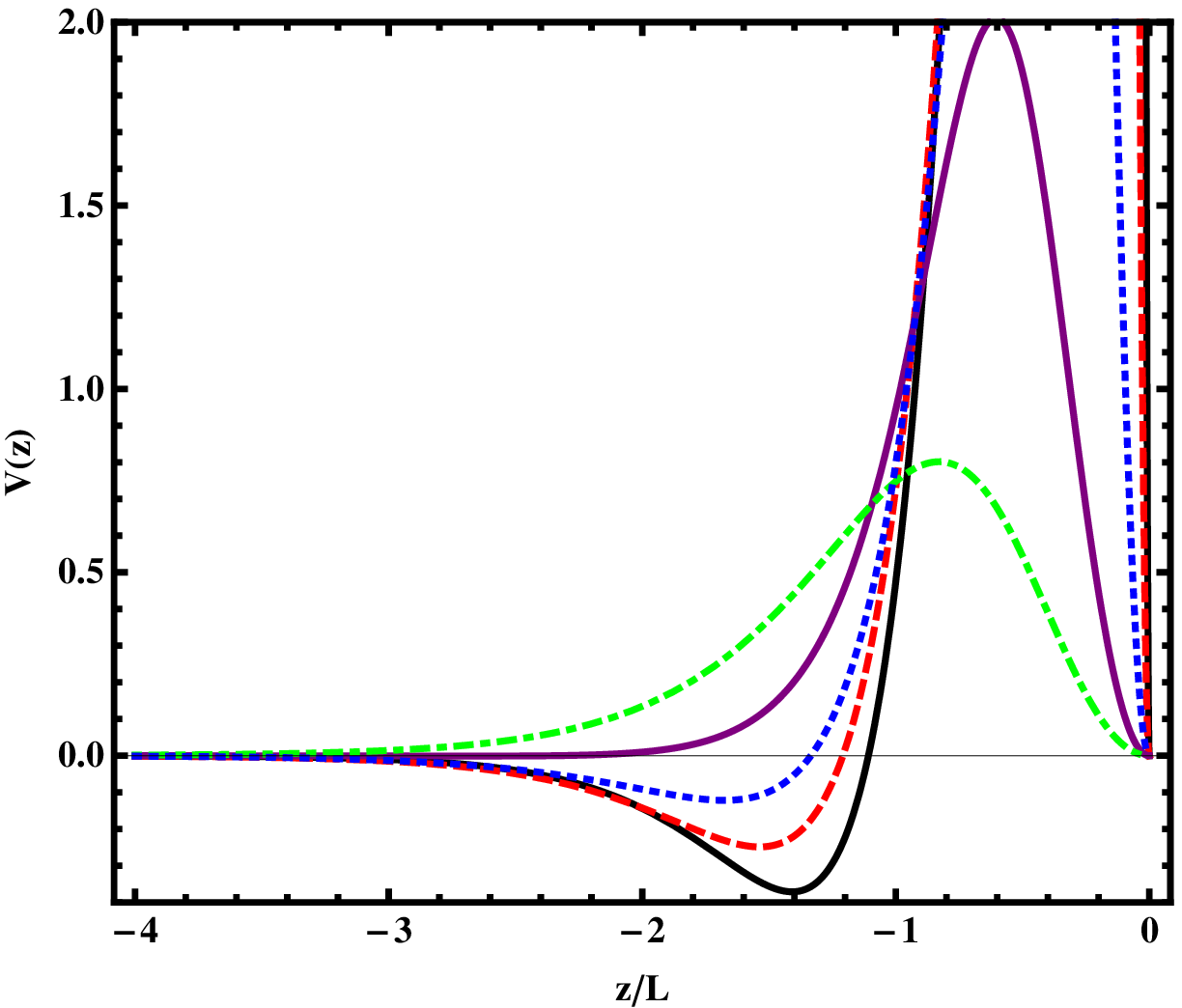,width=7cm,angle=0}
\end{tabular}
\caption{Effective potential in eq.~(\ref{schroedinger}).
Left: different models at fixed $T/T_c\sim0.3$. Black lines
refer to the case $b=0$ and $a=3$ (straight line) and
$a=\sqrt{3}$ (dashed line). Red lines refer to $b=1/\sqrt{3}$
and $a=\sqrt{3}$ (straight line) and $a=1/\sqrt{3}$ (dashed line).
According to eq.~(\ref{critical_a}), straight and dashed lines correspond
to potentials approaching $z=-\infty$ from below and above respectively,
as shown in the inset. Right: the model $a=\sqrt{3}$ and $b=1/\sqrt{3}$
for different temperature. From below to top
$T/T_c\sim0.2,\,0.3,\,0.5,\,,0.8,\,1$. The lower the temperature,
the deeper the minimum.\label{fig:pot_schroe}}
\end{center}
\end{figure}

The other distinctive feature of the behavior of $\s(\w)$, the
presence of a peak at small frequencies when $T > 0$, is due to the
fact that the potential of the Schr\"odinger problem is not positive
definite and can support a resonance near $\omega = 0$, causing a
sharp increase in the DC conductivity. Using the near-extremal
solution (\ref{nem}) we can clarify when to expect a peak in $\s(0)$
and when this peak will be present even at very low temperatures.
The explicit form of the Schr\"odinger potential near the horizon is
in this case \be V(z) \sim \frac{A^2}{r_0} \left [ 2 - 2h - \left (
b + \frac{a}{2} \right ) \xi \right ] e^{Az} \ , \ee where $A = \l_0
\h r_0^{1-b \xi}$. The potential, as already observed, receives a
negative contribution from the term due to the non-minimal coupling
between the scalar and the gauge field and a positive contribution
from the term due to the backreaction of the metric. From the
previous expression we can see that the potential approaches zero
with a negative exponential tail as $z \to - \infty$ if $a^2 + 2b^2
+ 3ab
> 4$. For a given $b$ this happens for
\be a > \frac{1}{2}\left [ \sqrt{b^2+16} - 3 b \right] \ .
\label{critical_a}\ee In particular when $b=0$ this implies that the
potential for the quasi-extreme black hole is positive near the
horizon for $a < 2$ and negative for $a > 2$. In Figure
\ref{fig:pot_schroe} we display the potential for selected values of
the temperature and the parameters to confirm that its behavior
changes according to Eq.~$(\ref{critical_a})$. When a negative
exponential tail is present for the near-extremal solution, the
value of the conductivity increases significantly at small
frequencies even at low temperatures, as shown in Fig.~\ref{fig:conductivityBR2}.
One can observe a small increase in
$\s(\w)$ at low frequencies even when the near-extremal potential is
positive. This is due to the fact that at temperatures higher than
those corresponding to the near-extremal black hole the term due to
the non-minimal coupling can dominate the term due to the
backreaction.

%
A similar increase in $\s(\w)$ at low frequencies was recently
observed in models where the Born-Infeld action of a probe brane is
coupled to a geometry with a Lifshitz scaling symmetry
\cite{Hartnoll:2009ns}.

\begin{figure}[ht]
\begin{center}
\begin{tabular}{cc}
\epsfig{file=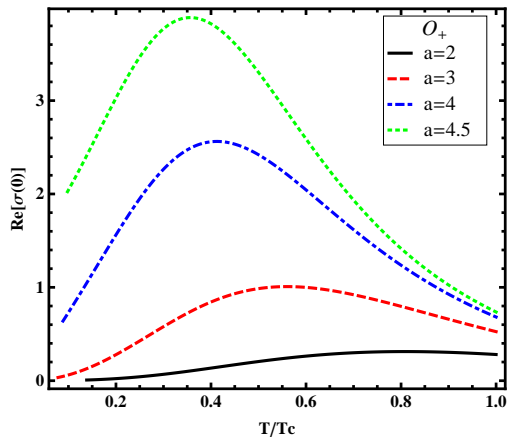,width=7cm,angle=0}
\end{tabular}
\caption{DC conductivity for
$V=-\frac{6}{L^2}\cosh(\phi/\sqrt{3})$ and $f=\cosh(a\phi)$.
From bottom to top $a=2,\,3,\,4,\,4.5$.\label{fig:DC_cond}}
\end{center}
\end{figure}

Another interesting feature of our model is that the DC conductivity $\s(0)$
depends in a non-monotonic way on the temperature, as shown
in Fig.~\ref{fig:DC_cond}. This effect becomes more evident as the value of $a$ increases.
In terms of the resistivity Fig.~\ref{fig:DC_cond} shows that there is a minimum
at low temperature. A minimum in resistivity is observed in metals containing magnetic
impurities and its presence was explained by Kondo as resulting from
the interaction between the magnetic moment of the conduction electrons
and the impurity. It would be interesting to identify in our model
what kind of effective interaction is induced among the charge carriers
by the scalar condensate which causes the minimum in resistivity.

%

\section{Conclusions}\label{sec:conclusions}
We have investigate  a broad  class of Einstein-Maxwell-dilaton
gravity  models in 4D AdS spacetime which admit both AdS-RN and
charged dilatonic black hole solutions. Below a critical temperature
the AdS-RN solution is unstable and it undergoes a phase transition
whose endpoint is the charged black hole dressed with a scalar
field.

This instability is interesting from a purely gravitational perspective
but it acquires a particular relevance when used in the context of the AdS/CFT
correspondence to provide a holographic description
of condensed matter phenomena. Our model describes a second order phase transition in the dual field
theory in which a neutral scalar operator condenses below a critical temperature.

The new phase has interesting electric transport properties,
presumably caused by the interactions of the charge carriers with
the scalar condensate. When the temperature is not too close to
zero, the optical conductivity has a minimum at low frequencies and
then a ``Drude peak'', reaching a constant value at $\omega=0$ which
can be considerably larger than its constant value at high
frequency. This effect is particularly evident for large values of
the non-minimal coupling between the scalar and the gauge field.
Another very interesting feature is that the resistivity does not
increase monotonically with the temperature but displays a minimum.
This effect is reminiscent of the Kondo effect, caused in real metal
with magnetic impurities by the interactions of the magnetic moment
of the conduction electrons with the magnetic moment of the
impurity. It would be interesting to compute the precise temperature
dependence of the resistivity and to clarify what kind of effective
interaction causes the minimum in resistivity.

We also studied the extremal limit of the charged dilatonic black
hole. The near-horizon metric has a Lifshitz scaling symmetry which
is a simple generalization of the one found in
\cite{Goldstein:2009cv}. Using the extremal solution we clarified
the behavior of the numerical solutions in the zero temperature
limit.

The analysis of the fluctuations of fermionic fields in the AdS-RN
\cite{Rey:2008zz, Lee:2008xf, Liu:2009dm, Cubrovic:2009ye,
Faulkner:2009wj} and in the charged dilaton black hole background
\cite{Gubser:2009qt} provides evidence for the existence of a Fermi
surface in the boundary theory. While in the former case the
macroscopic ground state entropy prevents a simple interpretation of
the dual phase, in the latter case one observes a more conventional
behavior, with vanishing zero-point entropy and with a specific heat
linear in the temperature. Since the phase transition discussed in
this paper connects the AdS-RN and the charged dilatonic black hole,
it may help to clarify the correct interpretation of the
corresponding dual theories.

It would be interesting to study models which allow for both the
instability studied in this paper and the instability caused by the
minimal coupling of a charged scalar field. One could then
investigate the effect of the neutral condensate on the properties
of the superconducting phase. Our models can in fact be easily
adapted to describe holographic superconductors by interpreting the
scalar field as the modulus of a complex scalar and adding to the
Lagrangian a coupling between the phase of the scalar and the gauge
field. Models of this type were recently discussed in
\cite{Aprile:2009ai}.

In order to identify precisely the dual theory and specify its
operator content and Hamiltonian, it would be interesting to embed
these black hole solutions
into ten or eleven dimensional supergravity.
For the holographic superconductors examples of this embedding were discussed in
 \cite{Gauntlett:2009dn, Gubser:2009gp, Gubser:2009qm,
Gauntlett:2009bh}.

\appendix
\section{Scalar hairs in the probe limit. Schwarzschild-AdS background}\label{app:probe_limit}

\begin{figure}[ht]
\begin{center}
\begin{tabular}{cc}
\epsfig{file=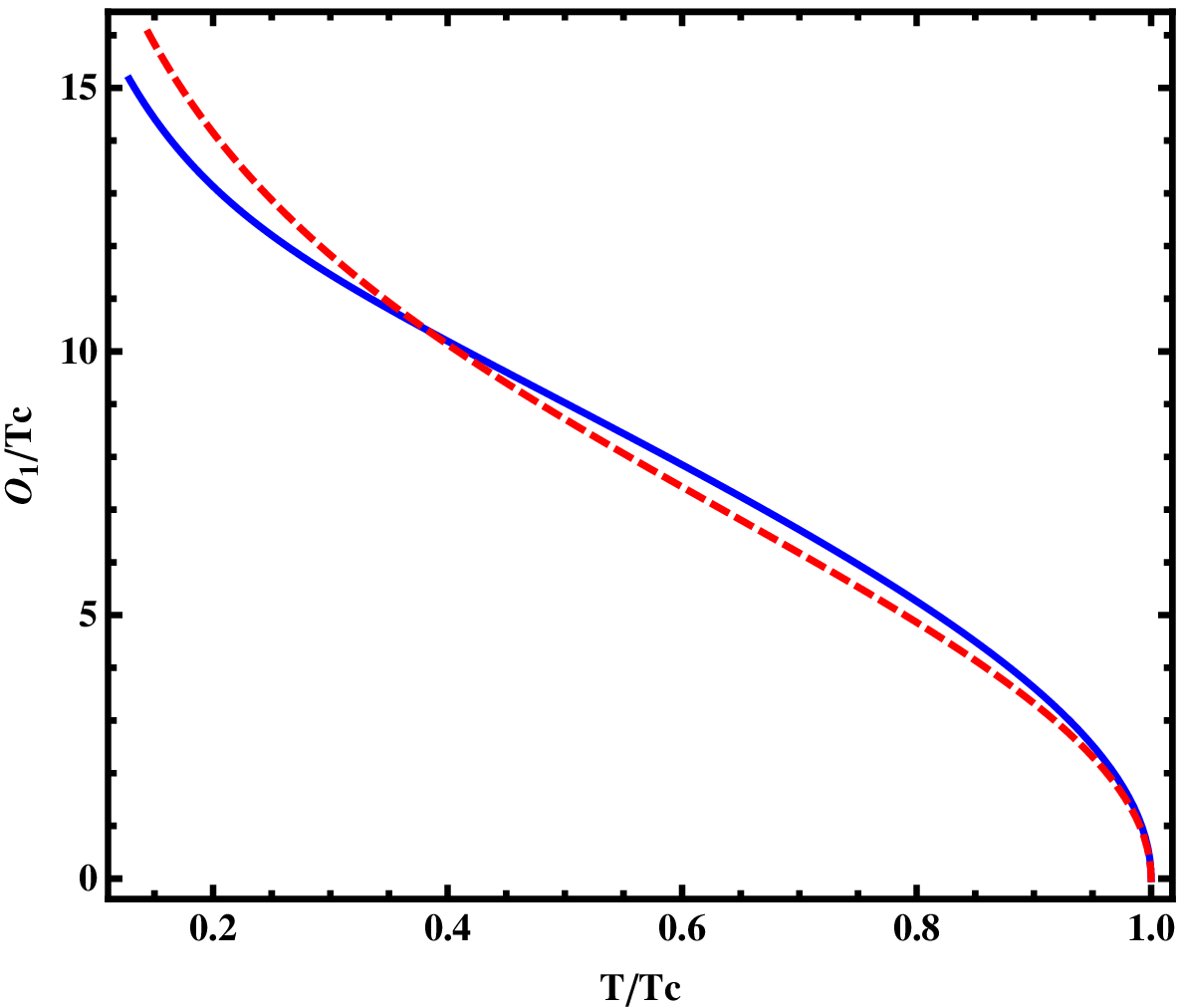,width=7cm,angle=0}&
\epsfig{file=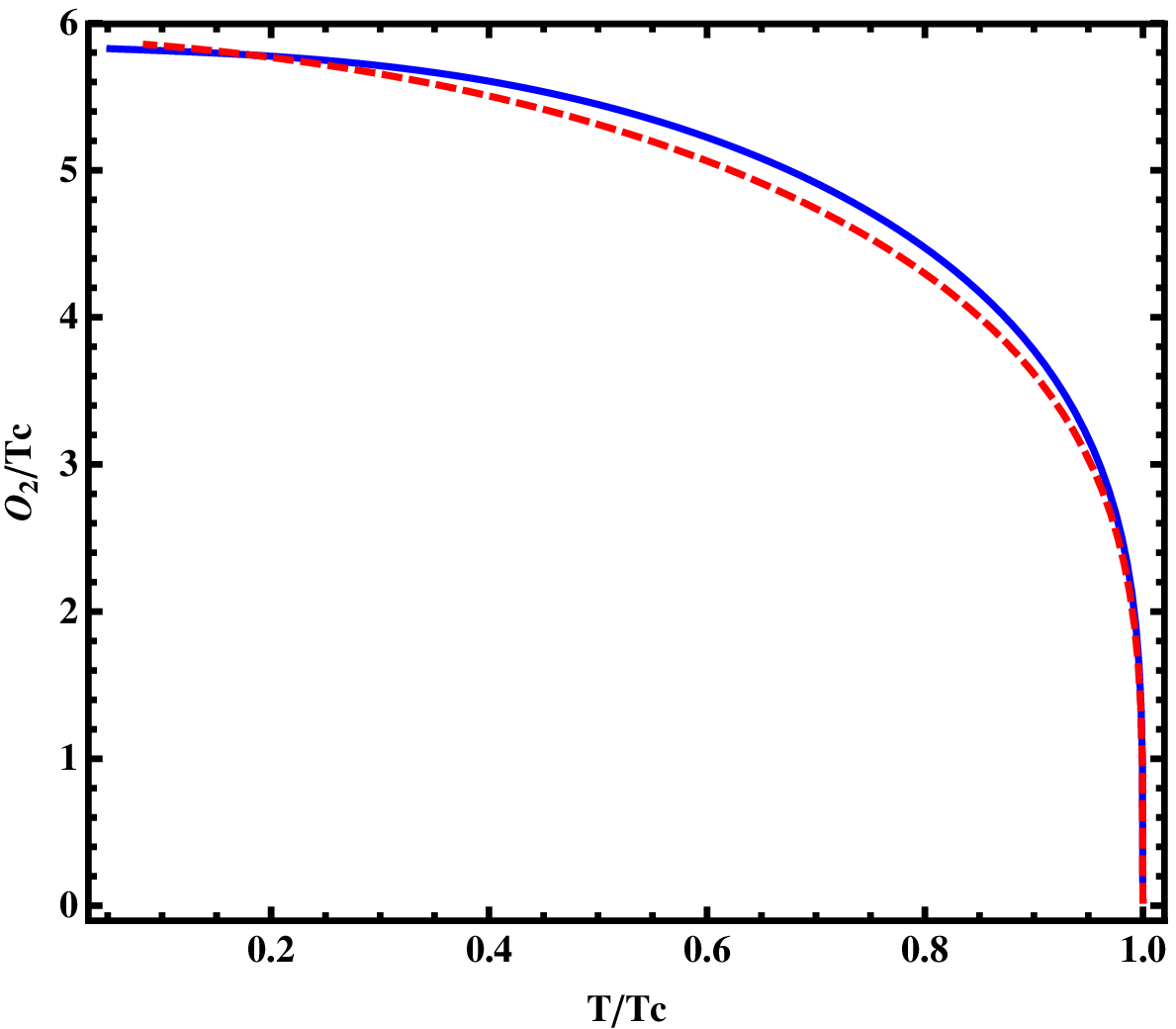,width=7cm,angle=0}
\end{tabular}
\caption{Condensate of the scalar operator as a function of the
temperature for ${\cal O}_-$ (left) e ${\cal O}_+$ (right). Straight
blue line is $f(\phi)=\cosh(\phi)$ and dashed red line is
$f(\phi)=1+\phi^2/2$. In both cases the potential is $L^2 V(\phi) =
- 6 - \phi^2$. For the two operators we have
$T_c\sim0.182\sqrt{\rho}$ and $T_c\sim0.135\sqrt{\rho}$
respectively. \label{fig:condensate}}
\end{center}
\end{figure}

Although in Sect. \ref{sec:backreaction} we derived numerically the
charged dilatonic black hole solution, it is of some interest to consider the
regime in which one can neglect the back reaction of the matter fields on the gravitational field.
The results of this Appendix can be compared with similar results obtained for
the holographic superconductors in Ref.\cite{Hartnoll:2008vx}.
We shall follow closely the methods used in that paper,
to which we refer  for further details.
We focus on potential of the form
$V(\phi)=-{6}/{L^2}-{\phi^2}/{L^2}$. The outcome does
not qualitatively change if one chooses a different $V$ as long as Eq.~(\ref{expVandf}) is satisfied.
The limit in which the dynamics for the scalar and
electromagnetic field  decouples from the gravitational dynamics is obtained for
$\alpha\to\infty$, after
the rescaling $\phi\to\phi/\alpha$ and $A_0\to A_0/\alpha$.
The gravitational part of the dynamics, represented by the sourceless
Einstein equations, is solved by
the planar AdS-Schwarzschild (AdS-S) background
\be
ds^2=-g(r)dt^2+\frac{dr^2}{g(r)}+r^2(dx^2+dy^2)\,,\label{bg:SAdS}
\ee
with $g(r)=r^2/L^2-2M/r$. The black hole  horizon is located  at
$r_h=(2ML^2)^{1/3}$, and the temperature is given  by
$T=\frac{3(2M)^{1/3}}{4\pi L^{4/3}}.$ After using
eq.~(\ref{cond:gaugefield}) the dynamics for  the scalar field is
described by a single equation
\be
\phi''(r)+\left[\frac{2}{r}+\frac{g'}{g}\right]\phi'(r)+\frac{1}{g}\left(\frac{\rho^2}{2r^4}\frac{1}{f(\phi)^2}
\frac{d f}{d \phi} - \frac{d V}{d \phi} \right)=0\,, \ee 
which has to be numerically integrated. The numerical integration is
performed using the method explained in Sect.
\ref{sec:backreaction}. Setting either ${\cal O}_-=0$ or ${\cal
O}_+=0$ in eq.~(\ref{bc_scalar}), we find a one-parameter family of
solutions. Varying the single free parameter $\phi_h$ we obtain the
condensate as a function of the temperature $T/T_c$. Numerical
results are shown in Figure \ref{fig:condensate} and they represent
the $\alpha\gg1$ limit of the exact solution described in the main
text.
We plot the results for  two different choices of the
coupling functions: $f(\phi)=\cosh(\phi)$ and $f(\phi)=1+\phi^2/2$.
The
condensate ${\cal O}_-$ diverges as $T\to0$. The divergence is an artifact of the
probe limit approximation and it is removed once the backreaction is taken into account, as shown in Figure
\ref{fig:condensate_backreaction}.

\begin{figure}[ht]
\begin{center}
\begin{tabular}{cc}
\epsfig{file=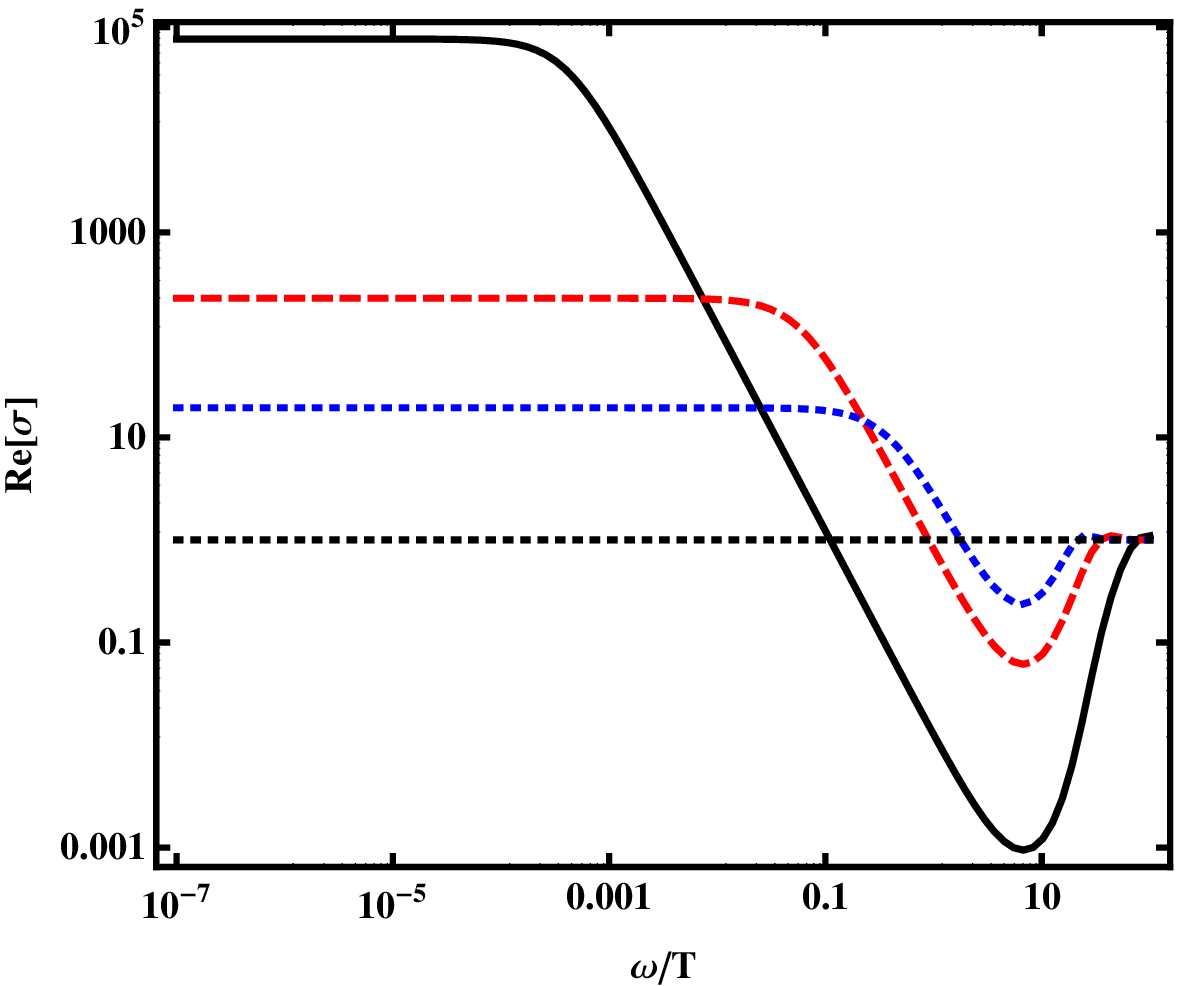,width=7cm,angle=0}&
\epsfig{file=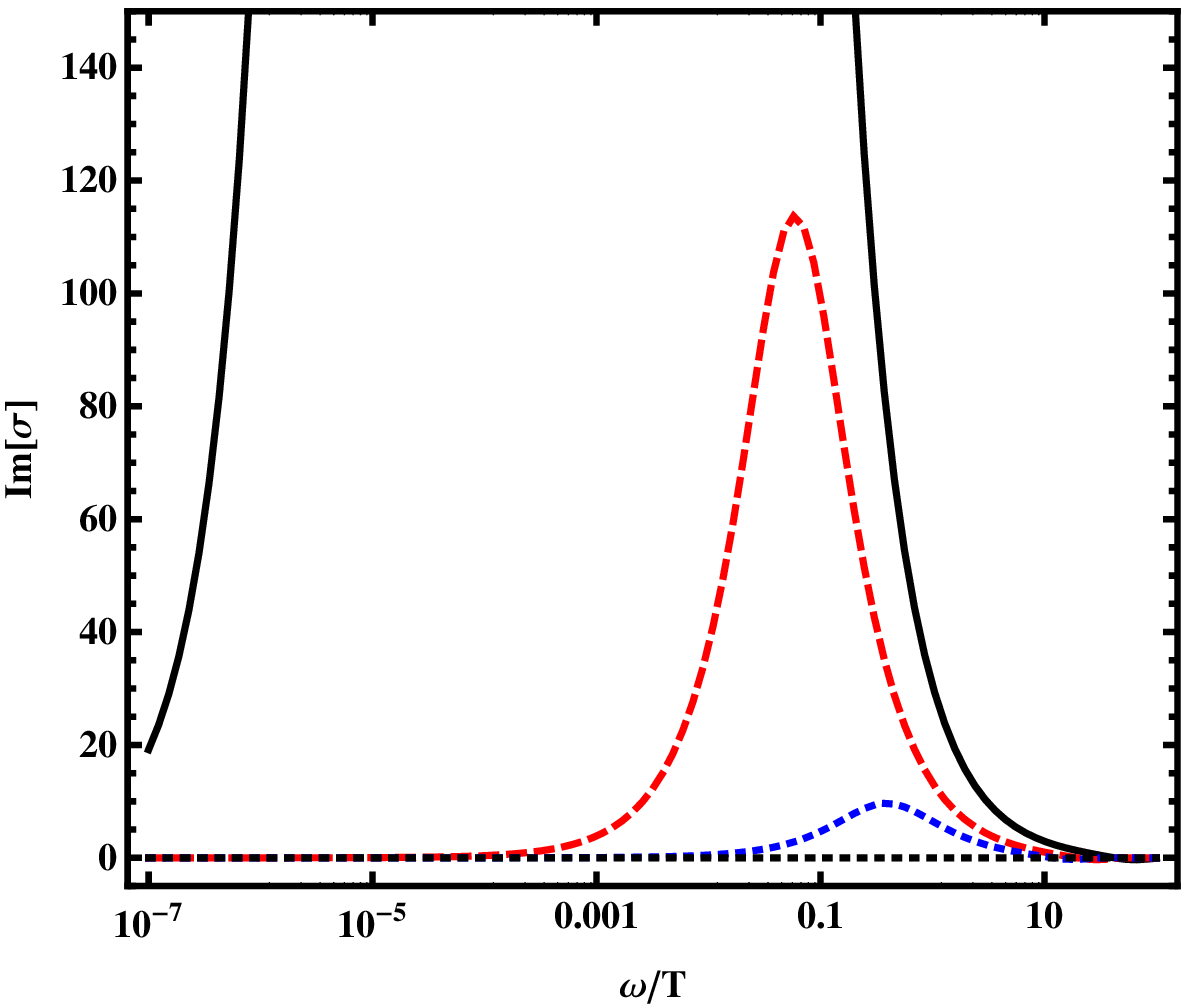,width=7cm,angle=0}
\end{tabular}
\caption{Real (left) and imaginary (right) part of the conductivity in a log scale for ${\cal  O}_+$
and $f(\phi)=\cosh(\phi)$. $T/T_c\sim0.14\,,0.3\,,0.5\,,1$ from bottom to top.
\label{fig:conductivity_bis}}
\end{center}
\end{figure}

We conclude with a brief discussion of the conductivity in the probe limit.
The perturbation equation reads
\be
A_x''(r)+\left(\frac{g'}{g}+\frac{1}{f(\phi)}\frac{df(\phi)}{d\phi}\phi'\right)A_x'(r)+\frac{\omega^2}{g^2}
A_x(r)=0\,.\label{pert_maxwell2}
\ee
Using the AdS/CFT definition (\ref{definition_sigma}) we obtain the
results shown in Figure \ref{fig:conductivity_bis} for the real and
imaginary part of the conductivity and for the condensate ${\cal
O}_+$. Similar results are obtained for ${\cal O}_-$. The increase
of $\s(\w)$ at low frequencies, typical of the dilaton black holes
considered in this paper, is evident in the probe limit. In fact it
grows exponentially as $T/T_c\to0$ but this is an artifact of the
probe limit and disappears once the back reaction is taken into
account (see Fig.~\ref{fig:conductivityBR1}).
%
Finally, since translation invariance is broken by the AdS-S background, the imaginary part does
not have a pole but vanishes for $\omega\to0$. From the dispersion relations it follows
that the DC conductivity is finite at $\omega=0$ in the probe limit.

\bibliography{phasetransrev}
\end{document}